\documentclass[pra,notitlepage,twocolumn,superscriptaddress,longbibliography]{revtex4-1}

\usepackage{amsmath,amssymb,amsfonts,bbm,graphicx,times}
\usepackage[dvipsnames]{xcolor}
\usepackage{braket}
\usepackage[normalem]{ulem}
\usepackage[colorlinks=true,bookmarks=false,linkcolor=RoyalBlue,urlcolor=RoyalBlue,citecolor=RoyalBlue,breaklinks]{hyperref}
\usepackage{subfigure}
\usepackage[capitalize]{cleveref}

\usepackage{booktabs,multirow}
\usepackage{pgfplots}
\usetikzlibrary{calc,patterns,angles,quotes}
\newcommand{\ketbra}[2]{\ensuremath{| \: #1 \:\rangle \hspace{-2pt} \langle \: #2 \:  |}}

\renewcommand{\phi}{\varphi}
\renewcommand{\theta}{\vartheta}
\renewcommand{\epsilon}{\varepsilon}
\renewcommand{\rho}{\varrho}
\newcommand{\unimi}{Quantum  Technology  Lab, Dipartimento  di  Fisica  ``Aldo  Pontremoli'',  Universit\`{a}  degli  Studi  di  Milano, I-20133  Milano, Italy}
\newcommand{\infn}{Istituto Nazionale di Fisica Nucleare, Sezione di Milano, 
I-20133  Milano, Italy}
\newcommand{\polimi}{Dipartimento di Fisica, Politecnico di Milano, I-20133  Milano, Italy}
\newcommand{\cnr}{Istituto di Fotonica e Nanotecnologie, Consiglio Nazionale delle Ricerche, I-20133  Milano, Italy}
\begin{document}
\title{Quantum spatial search in two-dimensional waveguide arrays}
%\author{C. Benedetti, D. Tamascelli, M.~G.~A. Paris, and A. Crespi}
\author{Claudia Benedetti}
\affiliation{\unimi}
\author{Dario Tamascelli}
\affiliation{\unimi}
\author{Matteo G.~A. Paris}
\affiliation{\unimi}
\affiliation{\infn}
\author{Andrea Crespi}
\affiliation{\polimi}
\affiliation{\cnr}

\date{\today}

\begin{abstract}
Continuous-time quantum walks (CTQW) have shown the capability to perform efficiently the spatial search of a marked site on many kinds of graphs. However, most of such graphs are hard to realize in an experimental setting. Here we study CTQW spatial search on a planar triangular lattice by means of both numerical simulations and experiments. The experiments are performed using two-dimensional waveguide arrays fabricated by femtosecond laser pulses, illuminated by coherent light. We show that the retrieval of the marked site by the quantum walker is accomplished with higher probability than the classical counterpart, in a convenient time window placed early in the evolution.
\end{abstract}
\maketitle

\section{Introduction}

The goal of spatial search algorithms is to find a marked item in a collection of 
entries, or a database, in the presence of displacement constraints. Such constraints 
define a topology, i.e. which of the entries are directly connected, and can be modelled as the edges of a graph $G$, whose vertices are the entries of the database. One simple 
strategy to find the target item is to traverse the graph along its edges, upon defining 
a classical random walk on the graph vertices. Such strategy would result in an average hitting time, i.e. number of jumps needed to hit the target item, linearly scaling with the number of vertices $N$ of $G$ \cite{chen08}. 

Quantum random walks on graphs exhibit, in general, a different behaviour than their classical counterparts \cite{kempe003,Venegas12,childs02,gualtieri20}. A continuous-time quantum walk (CTQW), for example, spreads on the line quadratically faster than its classical analogue, whereas there are particular graphs, such as the glued-tree graph \cite{childs03} where such ``speed-up'' is even exponential. Because of the close connection between graph traversing and the solution of decision problems, these observations have triggered an intensive research, aiming to assess the possible gain achievable by quantum walks, both in the continuous-\cite{farhi98,childs2004,childs09,tama13,tama14,tama16,chakra20,sym13010096} 
and in the discrete-time \cite{aharonov93,kempe03,shenvi03,lovett10,koch18} setting.
Spatial search on graphs by means of CTQWs has been thoroughly investigated in the last decades. Differently from a quantum transport setting, which focuses on moving a localized walker from one input site to a target one, in a spatial search the initial state of a search protocol is a uniform distribution over all the sites of the graph, indicating unbiasedness toward the target state. Moreover, an oracle is necessary in quantum search to correctly identify the solution node, while this is not required in transport where the target vertex is well identified. Starting from the seminal work from Childs and Goldstone \cite{childs2004}, the advantage provided by CTQW to quantum spatial search 
on different families of graphs has been analyzed \cite{childs14,meyer15,
philipp16,yasser16,osada18, glos19, osada20,portugal18}.

Implementing the spatial search algorithm in experiments is not trivial. To realize a CTQW, an experimental system has to be realized where a quantum particle can tunnel among the sites of a network, which are interconnected by some physical interaction.  However, the simplest versions of the spatial search algorithm based on CTQW 
%work efficiently 
{ finds the target with unit probability }
only for spatial dimensions equal or larger than 4 {(hypercube)}, or for fully-connected graphs \cite{childs2004}. Since physical interactions rapidly decay with the distance, it is hard to implement a physical system of quantum sites with such a high degree of connectivity. Variations of the original algorithm have been shown to work efficiently also for two-dimensional graphs with Dirac points in the spectrum \cite{foulger14, childs14}, such as honeycomb graphs. 
 Other algorithms, working in two-dimensional graphs such as triangular ones, require on the other hand discrete-time quantum walks \cite{tulsi08,abal12} and pose additional experimental difficulties, such as the implementation of the coin operator and of the conditional shift operator.
%To our knowledge, in fact,
{Whereas a large number of implementations of CTQW for the study of quantum information transfer and diffusion processes in different physical platforms are described in the literature \cite{manouchehri2006, Gr_fe_2016},} 
a single experiment of CTQW-based spatial search, implementing the spatial search on a {honeycomb lattice}, has been reported \cite{bohm15}. Such experiment used classical waves in a microwave setup containing weakly coupled dielectric resonators. In addition, the spatial search algorithm was recently run on a photonic processor \cite{Qiang21}; however, in that case the Hamiltonian was artificially codified in a multi-photon quantum interference experiment.

{In this paper,  we address the experimental implementation of 
quantum spatial search on two-dimensional triangular graphs by means 
of a photonic CTQW. In particular, graphs are encoded in two-dimensional arrays of optical waveguides fabricated by the femtosecond laser writing technique \cite{szameit2006, crespi2012, feng2016}, while walkers are photons from a coherent continous-waved laser beam.} {The coherent beam illuminates the input facet of the array, providing a delocalized input state. Optical waveguides, weakly coupled by evanescent-field interaction, represent quantum sites across which photons can tunnel at controlled amplitudes along propagation, and thus provide a convenient setting for implementing a CTQW \cite{perets2008, poulios2014, caruso2016, tang2018}. The photon distribution is retrieved by imaging the output facet of the array onto a CCD camera. }

We show that, in the presence of an engineered  ``defect'' marking the target database entry, the walker wave function tends to concentrate at the defect allowing for an identification of the target.  To quantify the advantage gained by using CTQW, we introduce and evaluate a suitable figure of merit that takes into account the fact that time is a resource to be used parsimoniously. In particular, we observe that such concentration process is faster than the classical random search on the same graph in a convenient time window.

\section{The model}
\begin{figure}
\centering
\includegraphics[scale=1]{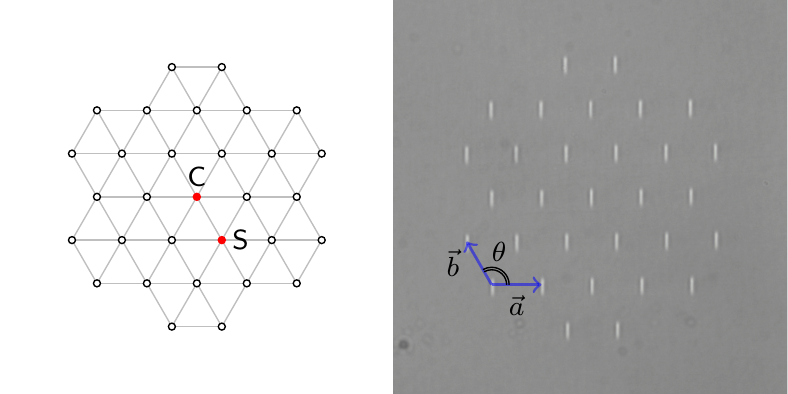}
\caption{\label{fig:geometry} The graph considered in this work is the finite portion of triangular lattice shown on the left, having $|V| = n = 31$ vertices. The target, if present, may be placed on the central site (C) or on a shifted position (S). On the right we report the microscope picture of the output facet of one waveguide array, implementing such graph in our experiments. The waveguide arrangement follows the Bravais parameters $\vec a$ and $\vec b$ with $|\vec a|=|\vec b|$ and $\theta=120^\circ$. For the array in this picture $|\vec a| = 23.4 \;\mu\mathrm{m}$ and the target is {placed} on the C site. Note that the target is implemented by a slight variation in the inscription parameters of the correspondent waveguide, which do not result in discernible changes in the appearence of the waveguide cross-section.}
\end{figure}

We consider the spatial search problem on finite triangular lattices, i.e. Bravais lattice with parameters $|\vec a| = |\vec b|, \theta = 120^\circ$ (see Fig.~\ref{fig:geometry}). The corresponding graph $G = (V,E)$ is fully defined by the $n \times n $ adjacency matrix $A$, where $A_{i,j} = 1$ if $\{i,j\} \in E$ and $0$ otherwise, and $|V|=n$ is the number of vertices. 

The basis state $\ket{j}$, $j=1,2,...,n$ of the $n$-dimensional Hilbert space of states $\mathcal{H}$,  identifies a configuration where a single excitation, or walker, is located at the vertex $j$. The adjacency matrix generates a quantum walk on $G$ through the unitary evolution ($\hbar=1$):
$    \ket{\psi(t)}  = e^{-i \gamma A t} \ket{\psi(0)}$,
with $\ket{\psi(0)}$ any initial state in $\mathcal{H}$ and $\gamma >0$ a constant setting the transition rate of the walker. In order to perform the search of a target  vertex $w$, we introduce a local perturbation in the form of a diagonal matrix $\beta \ketbra{w}{w}$, with $\beta >0$, which acts as an oracle \cite{childs2004}. The quantum Hamiltonian for the search of a target site $w$ reads, therefore
\begin{equation}
H^Q_w = -\gamma A -\beta \ketbra{w}{w}.
\label{hamQ}
\end{equation}
We will moreover consider the linear superposition of all the nodes as the initial condition $\ket{\psi_0} = \frac{1}{\sqrt{n}} \sum_{j=1}^n\ket{j}$.
The probability of finding the walker at the target site $w$ at time $t$,
\begin{equation}
    p_w^Q(t) = \left |\bra{w} e^{-i t H^Q_w} \ket{\psi_0} \right |^2.
    \label{probQ}
\end{equation}
We note that the parameter $\gamma$ governs the overall time-scale of the system dynamics. If one considers a normalized evolution time $\gamma t$, the probability $p_w^Q(\gamma t)$ will be determined by the ratio $\beta/\gamma$.
%%%%%%%%%%%%%%%%%%%%%%%%%%%%%%%%%%%%%%%%%%%%%%%%
%%%%%%%%%%%%%%%%%%%%%%%%%%%%%%%%%%%%%%%%%%%%%%%%
%
In order to compare the features of the quantum spatial search with those of a classical random search, we define the following  stochastic process on $G$. The dynamics of a classical random walk in the absence of the marked vertex is generated by the graph Laplacian matrix, defined as $L = A-D$, where $D = (d_1,d_2,\ldots, d_n)$ is a diagonal matrix and $d_j$ indicates the degree of the $j$-th vertex. 

In order to mark a site and to model the fact that when the process hits the target vertex $w$ the search task is completed, we modify the graph by making all the transitions to the marked vertex irreversible. We do that by setting the edges entering into $w$ as directed.
It follows that the generator of the classical dynamics is determined by the Laplacian of a directed graph, $L_c$, such that $[L_c]_{jw}=0\, \forall j$  ($w$ is called a sink) and with diagonal entries fixed such that the sum of the elements of each column of $L_c$ is equal to zero, i.e.  $\sum_j [L_c]_{jk}=0\,\forall k$.
In this way, $L_c$ is  a proper generator of the trace-preserving semi-group $\{ C(t) = \exp{(-\gamma t L_c)}, t \geq 0\}$, with $\gamma > 0$ the transition rate. For the sake of uniformity, we call $\ket{p_0} = \ket{\psi_0}/\sqrt{n} =(p^0_1,p^0_2,\dots,p^0_n) $ the vector describing the initial probability distribution of the classical walker on the graph sites, with the property that $\sum_j p^0_j=1$. The probability of reaching the target at time $t$ is given by:
\begin{equation}
    p_w^C(t) = \bra{w} C(t)\ket{p_0}.
    \label{probC}
\end{equation}

\section{Simulation of spatial search by CTQW}\label{theosearch}
\begin{figure}
    \centering
    \includegraphics[width=.99\columnwidth ]{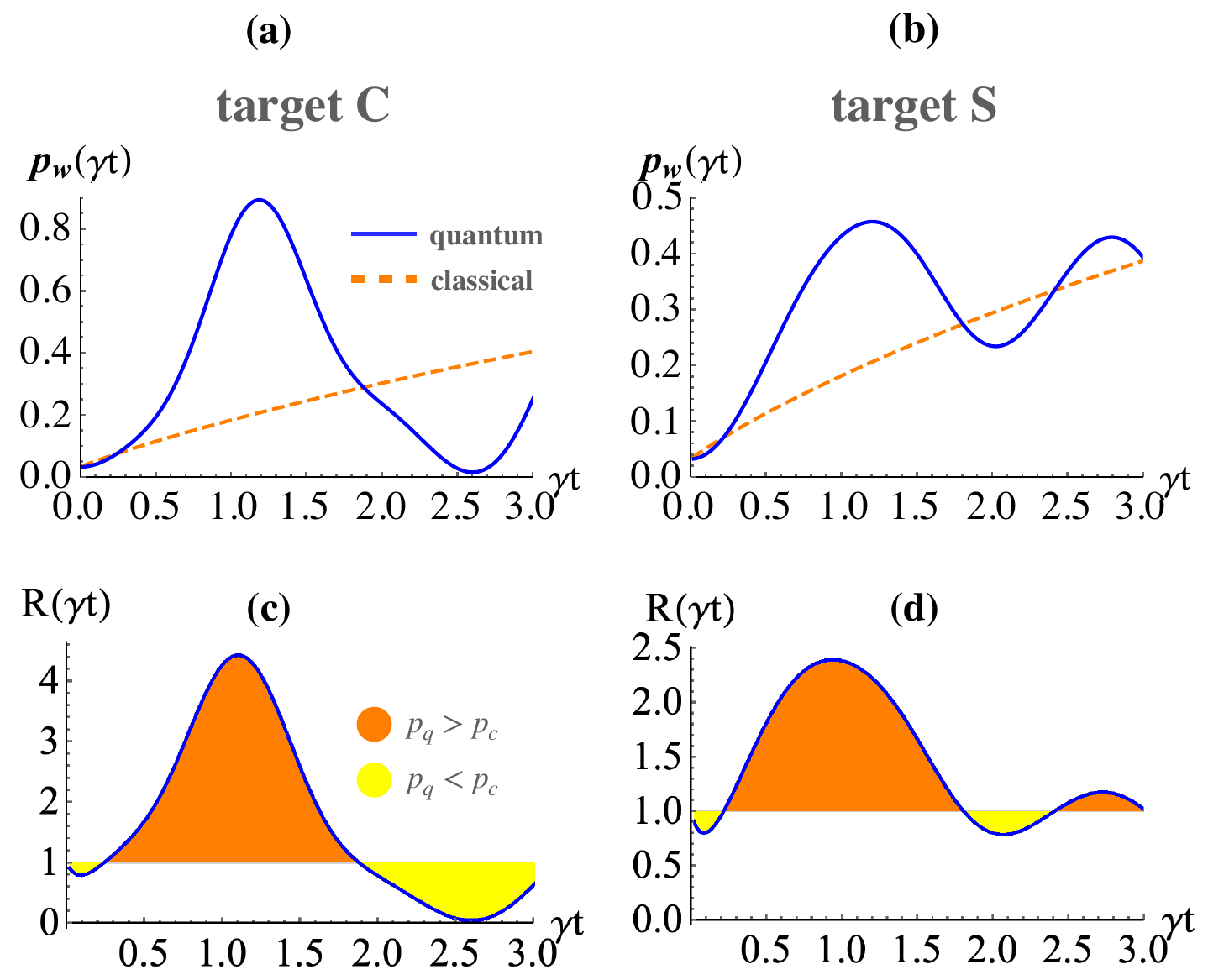}
    \caption{Time evolution of the probabilities $p_w(\gamma t)$ for the quantum (blue solid line) and the classical (dashed orange line) search, for the central (a) and shifted (b) target.
    In the lower panels we display the quantity $R(\gamma t)$ for the central (c) and shifted (d) target. As a guide for the eye, we color in orange the region where the quantum walk outperforms the classical walk, while the yellow shadow indicates the opposite regime. The detuning $\beta/\gamma$ is set to 4.16.
    }
    \label{fig:Rqc}
\end{figure}

We  study the spatial search  on   the finite graph sketched in Fig.~\ref{fig:geometry} composed of 31 sites, and we consider two possible positions for the target: the central site (C) and a site in a shifted position, adjacent to the central one (S). We report in Fig.~\ref{fig:Rqc}a,b the classical and quantum probability to reach the two different  target sites, calculated numerically through Eqs.~\eqref{probQ} and \eqref{probC} respectively. 

\begin{figure}[t]
    \centering
    \includegraphics[width=.99\columnwidth ]{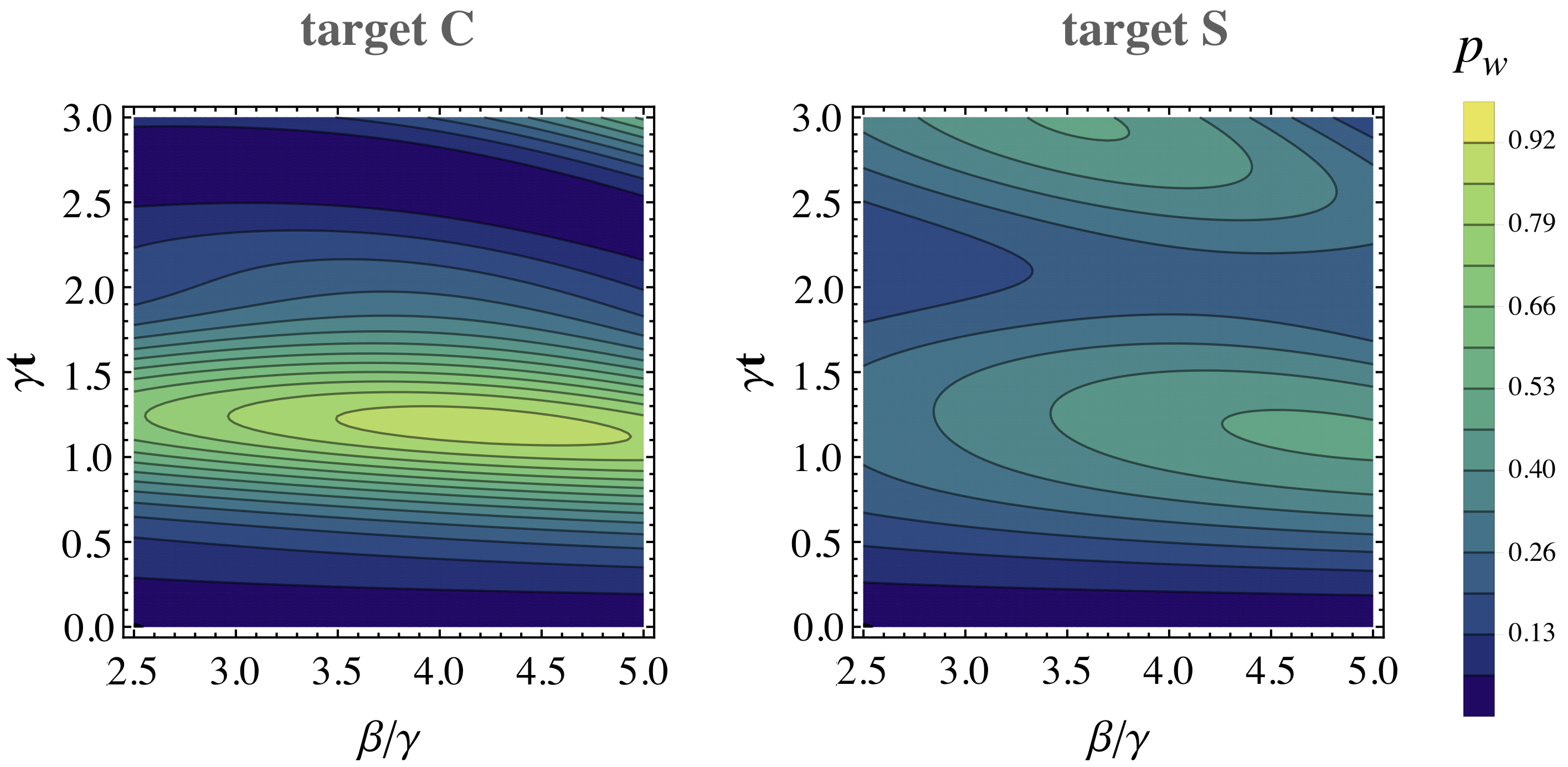}
    \caption{ Target probability $p_w^Q(\gamma t)$ as a function of the target onsite energy $\beta/\gamma$ and time $\gamma t$, for  central (C) and  shifted (S) target.}
    \label{fig:denbeta}
\end{figure}

In order to compute the quantum evolution, we fix in both cases $\beta/\gamma=4.16$, a value that optimizes the maximum value reached by $p^Q_w(\gamma t)$  for a target placed at the centre of the graph.  We thus make the target on-site energy independent of its location since, in principle, in a search problem the target location is not known a-priori. 
As shown in Fig.~\ref{fig:denbeta}, the qualitative behavior of the target probability $p_w(\gamma t, \beta/\gamma)$ presents a smooth dependence, for fixed value of the dimensionless time $\gamma t$, on the value of the {ratio} $\beta/\gamma$. It follows that a small detuning {in} $\beta/\gamma$ does not significantly degrade the spatial search. 

We note that $p^C_w(\gamma t)$ is monotonically increasing {in time}, because of the irreversible nature of the dynamics described by the generator $L_c$. Namely, in the classical search as defined in the previous section, the walker always finds the target node if we wait long enough. However, $p^Q_w(\gamma t)$ surpasses $p^C_w(\gamma t)$ in both considered cases, for wide time intervals early in the walk evolution.

To provide an immediate comparison of the quantum dynamics versus the classical counterpart, and thus reliable information about the speedup of the quantum search with respect to the classical case, we introduce the ratio $R(t)$ of the two probabilities:
\begin{equation}
    R(t)=\frac{p_w^Q(t)}{p_w^C(t)}.
   \label{resource}
\end{equation}
In fact, our aim is not only to identify the target node $\ket{w}$ using a CTQW, but also to perform the spatial search in the shortest possible time. Since in the classical case the probability of finding the target is a monotonically increasing function of time, while in the quantum case it has an oscillatory behaviour,  $R(t)$ naturally takes into account both the fact that we want the $p_W^Q(t)$ to be large and time to be small. Plots of $R(\gamma t)$, for the two considered target sites in our finite graph, are reported in {panels} (c) and (d) of Fig.~\ref{fig:Rqc}: The orange region highlights the time-intervals where the quantum walker outperforms its classical analogue in finding the target node. 
 Remarkably enough, we do not need to measure the walker position at a specific single time, but we have a large time interval in which the CTQW better identifies the target node with respect {to a classical continuous-time random walk}, i.e. $R(\gamma t) >1$.  

We should note that these results are not specific to the graph  of Fig.~\ref{fig:geometry}. The qualitative trend of the quantity $R(\gamma t)$ keeps indeed its validity if we increase the size of the graph. In Appendix~\ref{app:scaling} we report the scaling behavior of $R(\gamma t)$ for a triangular lattice with increasing number of vertices, and we %. We 
show that there is always an advantage in using a quantum walk to perform the search, with respect to its classical counterpart.


\section{Spatial search in a waveguide setting}\label{sec:spatialSearchWg}

In our experiments we study the spatial search problem in a photonic setting, exploiting lattices of single-mode waveguides, weakly coupled to each other. {Considering only first-neighbour coupling, the dynamics of one photon propagating in the array can be precisely mapped to a single-particle CTQW described by the Hamiltonian \eqref{hamQ}. In fact, the states $\ket{j}$ represent the occupation of the optical mode of the $j$-th waveguide of the array. Because of evanescent-field coupling, during the propagation the photon can tunnel in neighbouring waveguides with a rate $\gamma$. A detuning in the propagation constant of the $w$-th optical mode implements the parameter $\beta$. {In addition}, the evolution time is directly mapped onto the propagation coordinate; indeed, we can take directly the longitudinal spatial coordinate as a measure of the evolution time $t$.}

{We fabricated lattices of 31 waveguides in fused-silica substrates, with the geometry as in Fig.~\ref{fig:geometry}, using the femtosecond laser waveguide writing technique. In detail, we employed a commercial laser system (LightConversion Pharos) providing 180~fs pulses with up to 40~$\mu$J energy and repetition rates up to 1 MHz. For the waveguide inscription we used a repetition rate of 50~kHz and pulses with 250~nJ energy, which were focused in the bulk of the substrate by means of a 20$\times$ (0.45~NA) microscope objective. The average depth of the waveguide arrays below the substrate surface is 170~$\mu$m, for which the objective provides optimal compensation of spherical aberrations. 
Individual waveguides sustain a single guided mode at the wavelength of 633~nm, with mode size of about 13~$\mu$m ($1/e^2$ diameter) and propagation loss of about 0.5~dB/cm.}

\begin{figure*}[tbp]
\includegraphics[scale=0.9]{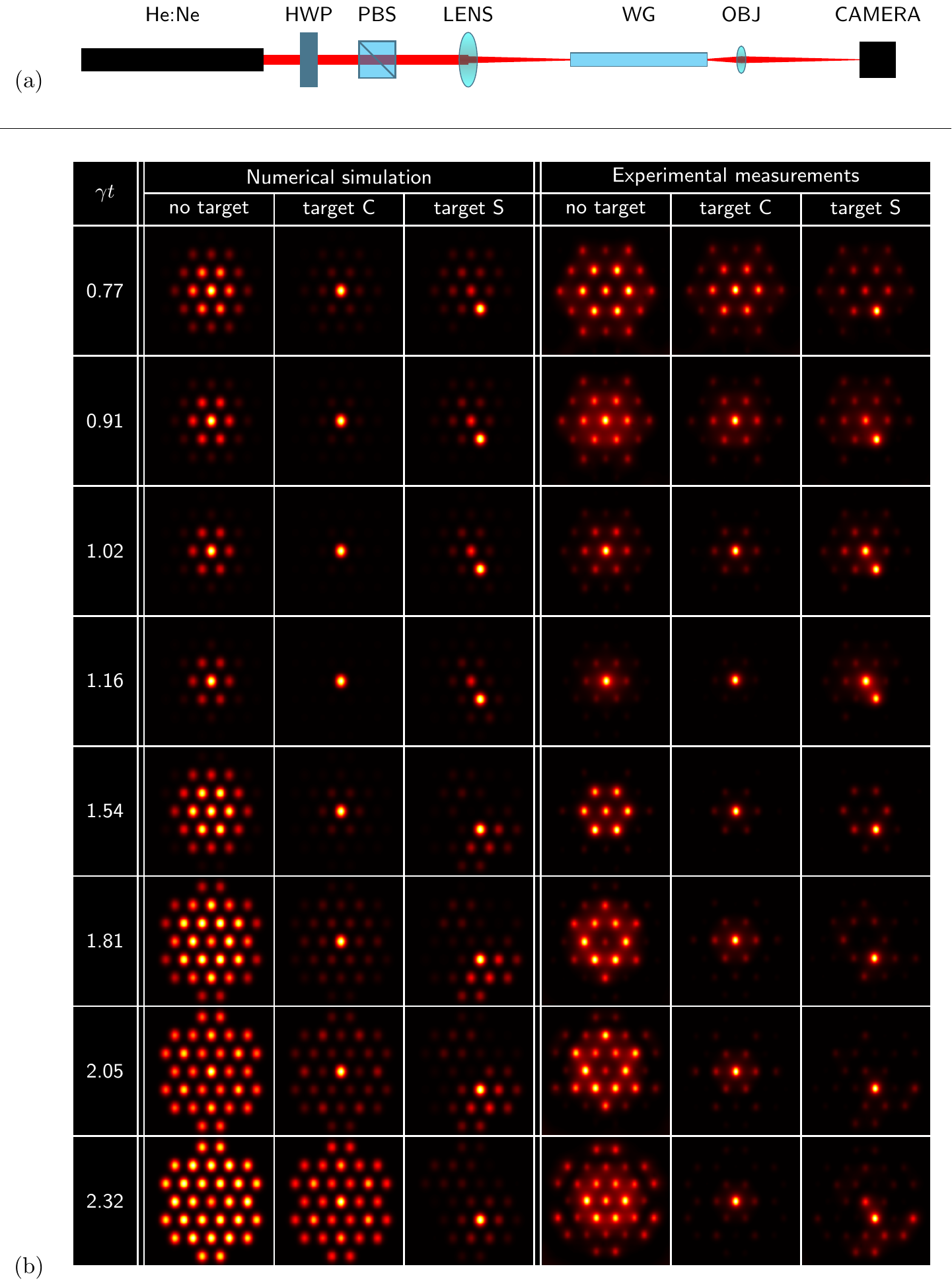}
\caption{\label{fig:exp} (a) Experimental apparatus used for the characterization of the waveguide arrays. A He:Ne laser beam (633 nm wavelength) is first passed through a half waveplate (HWP) and a polarizing beam-splitter cube, to arbitrarily attenuate the beam intensity and to select horizontal polarization state. The beam is focused by a convex lens (with focal length $f$~=~40~cm) on the input facet of the glass sample containing the waveguide arrays (WG). The glass sample is mounted onto micrometric manipulation stages allowing three orthogonal translations and two tilts. The output facet is imaged onto a CCD camera using a 10$\times$ objective. (b) On the right, the experimentally measured output distribution for the fabricated waveguide arrays, for the different values of nominal propagation coordinate $\gamma t$ and for the different target position (or absence of target). On the left, numerical simulation of the same distributions, assuming the nominal detuning $\beta /\gamma = 4.1$ on the target waveguide. The intensity scale of each picture is normalized to the maximum value of that picture.}
\end{figure*}

The waveguide arrays were realized with different length and waveguide separation, to provide different values of $\gamma t$ (see Appendix~\ref{sec:appFab}).  To implement the target site, the correspondent waveguide was inscribed with a different speed with respect to the other ones. In fact, for femtosecond laser written waveguides the coupling coefficient $\gamma$ is typically an exponentially decreasing function of the waveguide separation, while the propagation constant can be modulated by varying the writing speed \cite{crespi2012}. {We note that, for the inscription parameters adopted in our experiment, mode size and propagation losses are negligibly affected by these changes in writing speed.}
For a given value of $\gamma t$, three kind of arrays were fabricated, one next to the other in the glass: one without target (i.e. with all identical waveguides), one with the target on the site C, one with the target shifted on the site S (see Fig.~\ref{fig:geometry}). The nominal ratio $\beta/\gamma$ was kept fixed to  $ \approx 4.16$ well within the range of values that optimize the search of a target placed at the centre of the graph (see Fig.~\ref{fig:denbeta}). 

To realize the spatial search algorithm, we employed the apparatus sketched in Fig.~\ref{fig:exp}a. To provide a stream of photons at the input, delocalized with (approximately) equal probability amplitude and phase across all the optical modes of the array, %an input wavefront with (approximate) equal amplitude and phase on all waveguides, 
we mildly focused a He:Ne laser beam using a $f$~=~40~cm convex lens and we placed the input facet of the array in the focal point. 

The $f$~=~40~cm lens produces a focal spot of about 400~$\mu$m $1/e^2$ diameter, experimentally measured with the knife-edge technique, which is larger than the array size. {This beam diameter is sufficient to approximate, in our experiments, a uniform excitation; indeed, the output distribution is found experimentally to be marginally affected by small shifts in the transverse alignment of the beam with respect to the center of the array.} {Besides,} a Gaussian beam yields a planar wavefront in its waist and such planar {phase} condition is maintained with good approximation in the Rayleigh region, which in this case is a few centimeters long. Thus, {also} the longitudinal positioning of the glass chip is not critical. {On the other hand, the tilt degrees of freedom need to be carefully adjusted: if the propagation vector of the impinging beam is not parallel to the propagation vector of the optical modes of the array, a phase gradient results to be imposed to the input state, with possible severe consequences on the evolution (see Appendix~\ref{sec:appTilt} and  Fig.~\ref{fig:tilting}).}

Output light was collected with a 10~$\times$ (0.25~NA) microscope objective and imaged onto a monochromatic CCD camera. The pixel intensities are proportional to the average photon number impinging on the pixel area. We note that the output photon distribution measured by our experimental setup is identical to the one obtainable using pure single-photon sources at the input, and single-photon-sensitive imaging devices at the output.

\begin{figure}
\includegraphics[scale=1]{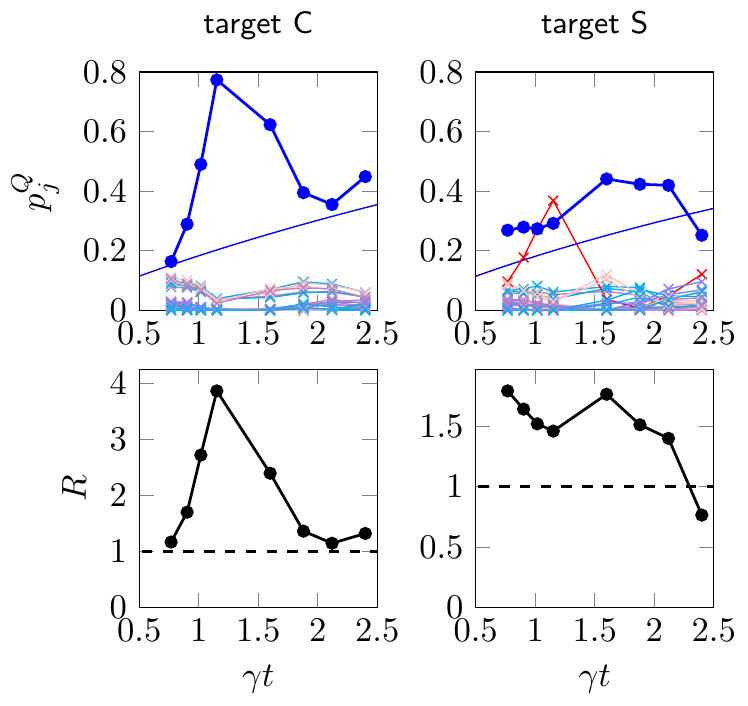}
\caption{\label{fig:population} Experimental values of $p^Q_j(\gamma t)$  at each waveguide output measured in the arrays with the target on the central site (top left graph) and in shifted position (top right graph). Blue circles represent the measured values for the target waveguide while crosses represent non-target sites. Blue thin lines represent the classical probability  $p^C_w(\gamma t)$ to find the target in a classical random walk setting. In the top-right graph (target S), red crosses correspond to the values of the central waveguide. 
Experimental values of $R(\gamma t)$ are plotted (bottom left and bottom right graphs), as calculated from the experimental values of  $p^Q_w(\gamma t)$ and the calculated values of $p^C_w(\gamma t)$, as plotted in the above graph.}
\end{figure}

\section{Experimental results}

Experimental images of the output light distribution, for all the fabricated waveguide arrays, are reported in Fig.~\ref{fig:exp}b. The figure also shows, for comparison, the results of numerical simulation, performed considering the nominal values for the normalized time $\gamma t$ and for the ratio $\beta/\gamma$ in each array. %\DT{Considering that the placement of the input facet of the glass chip provides equal probability amplitude and phase across all the optical modes of the array}, 
{Simulations also assumed an ideal flat input wavefront.} The simulated distribution of a classical random walker in analogous conditions is given, as a further comparison, in Fig.~\ref{fig:classical}.

Figure~\ref{fig:population} reports the occupation probability $p^Q_j(t)$ for each waveguide, obtained from the images in Fig.~\ref{fig:exp}b. In detail, to retrieve these values the pixel intensities were integrated within circles of about the size of the optical modes, centered on the position of each waveguide. The obtained distribution was then normalized to sum up to one. {We note that the optical losses of the waveguides, which are uniform across the array, do not influence the normalized distribution.}

From both figures we can observe a clear concentration of the light on the target waveguide (in the arrays {where a target is implemented}) for quite a wide range of values of $\gamma t$. {Importantly, in a large part of the investigated time interval, the probability to find a photon in the target waveguide is dominant over the correspondent probability for a classical walker to have reached that site. Thus the ratio $R(\gamma t)$, estimated by combining the result of the experimental measurements for the CTQW, and numerical simulations of the classical case, robustly exceeds one for both target positions.}

Some light {concentration} in the central waveguide is also observed in the case {of shifted target and even in the case} of the uniform arrays, {at evolution times $\gamma t < 1.5$ (see Fig.~\ref{fig:exp}). We attribute this to boundary effects, i.e. to the partial reflection of the delocalized initial state from the edges of the finite-sized lattice, which tends to focus in the middle at certain evolution times.} The occurrence of this spurious focusing effect{, which {influences} the convergence of the spatial search algorithm, explains why the target is reached with higher probability when this is placed } in the central position with respect to the case in which it is shifted.

The differences between the simulated and measured distributions, which are noted in Figs.~\ref{fig:exp}b and Fig.~\ref{fig:population}, may be due to contributions from several causes. {For instance,} the fabrication technique presents tolerances in implementing the desired values of coupling coefficients $\gamma$ and detuning $\beta$. In particular, the experimental laws that relate these quantities to the inter-waveguide distance and to the writing speed are crucially dependent on the microscopic refractive index distribution of the waveguide, which in turn is a result of complex interactions between the laser pulses and the substrate. Drifts in the performance of the femtosecond laser used for the fabrication (e.g. in the pulse length), may cause these laws to change slightly from day to day, with respect to the preliminary calibration. As a matter of fact, these non-idealities do not prevent the observation of a favourable localization of the quantum walker on the target, as discussed above.

\section{Conclusion and Outlook}
We have experimentally demonstrated the spatial search algorithm on a finite-size triangular graph, exploiting {two}-dimensional waveguide arrays and coherent light excitation. The target site was encoded as an engineered defect in the lattice. We have shown that there is a relevant time window in which the walker wavefunction concentrates at the target site with higher probability than the corresponding classical random search on the same graph.

{We have therefore provided evidence that 2D graphs, while not optimal, are able to achieve interesting results with a reasonable experimental effort. In particular, we have shown that the quantum advantage is clear also with these topologies, especially in comparison with a classical random walk search, without the need to implement a fully connected graph.}

Our results are particularly interesting in any situation where time is a constrained resource, and the database corresponds to a poorly connected structure. In those cases, in fact, finding the target with certainty is impossible with classical resources, and no simple quantum algorithms are available. Indeed, we have found that a simple CTQW may provide an advantage over the classical counterpart, in identifying the target  with larger probability. 

This study provides an experimental  implementation of the CTQW spatial-search algorithm in a photonic system, where the quantum walk dynamics is effectively reproduced in an analogical setting. Optical waveguide arrays are indeed a highly versatile platform, which may allow in the future to explore not only different graph geometries, such as square or hexagonal lattices, but also to implement different search protocols, as the vacancy-based search scheme proposed in \cite{foulger14}.

In addition, we observe that the search problem stated through Eq. \eqref{hamQ} can be reinterpreted as a strategy to identify defects in physical systems. The projector onto the target state $\beta \ketbra{w}{w}$ modifies the on-site energy of a particular vertex by a detuning $\beta$, being akin to the effect of a fabrication defect in a lattice. As shown in our numerical simulations, the localization effect is robust with respect to variations in $\beta$. Therefore, a CTQW which starts in a superposition of all sites may be used to detect the flawed node, i.e. the target site.

\section*{Acknowledgements}
DT has been supported by UniMi through the ``Sviluppo UniMi'' initiative.
AC acknowledges funding by the Ministero dell'Istruzione, dell'Universit\`a e della Ricerca (PRIN 2017 progamme, QUSHIP project - id. 2017SRNBRK). MGAP is member of INdAM-GNFM.
{The Authors would like to thank Dr.~Roberto Osellame for helpful discussions.}

%%%%%%%%%%%%%%%%%%%%%%%%%%%%%%%%%%%%%%%%%%%%%%%%%%%%%%%%%%%%%%%%%%%%%%%%%%%%%%%%%%%%
\appendix

\section{Scaling behaviour of the quantum search}\label{app:scaling}
\begin{figure}[t]
    \centering
    \vspace{0.5cm}
    \includegraphics[width=.99\columnwidth ]{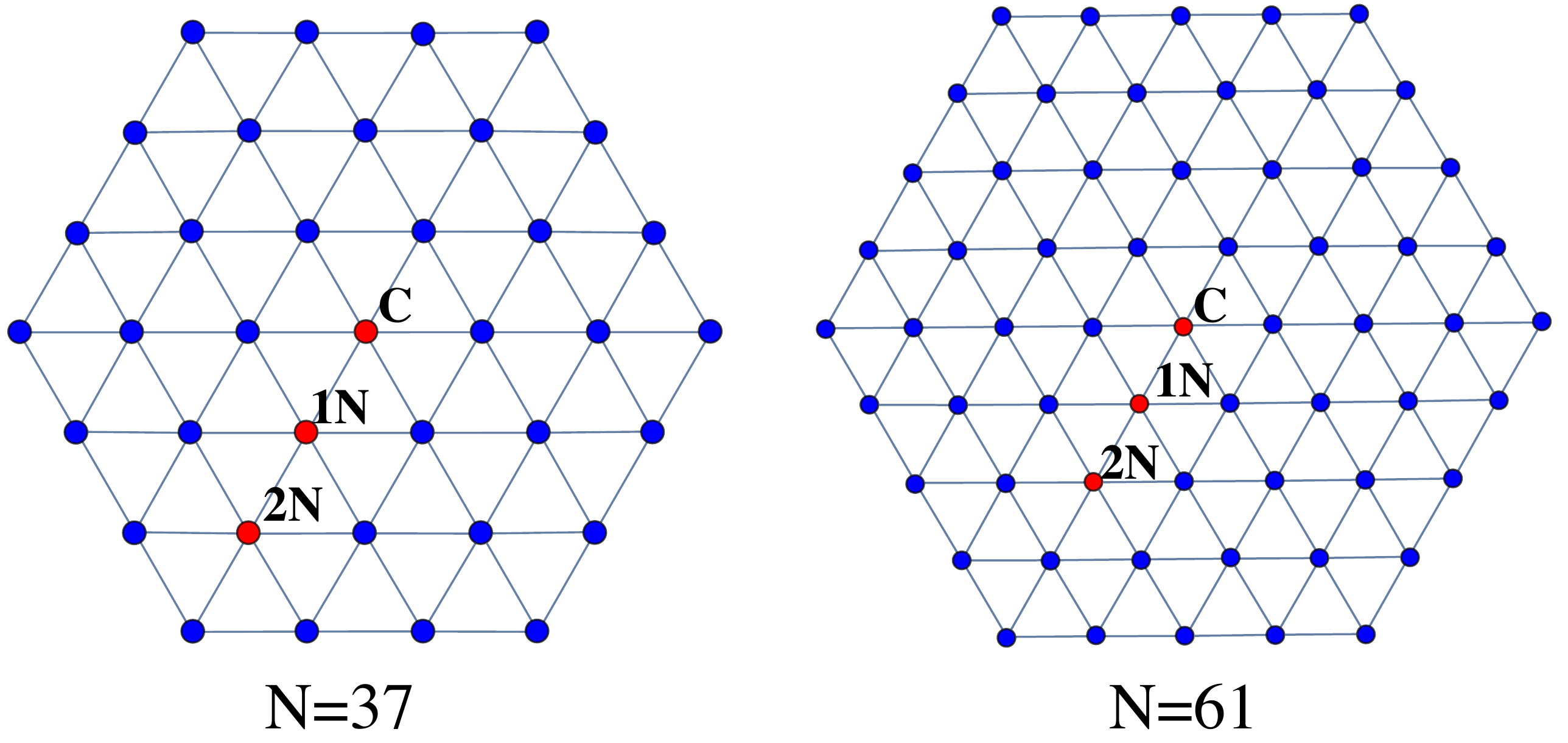}
    \caption{Triangular graph with $N=37$ (left) and $N=61$ (right) nodes. The central (C), nearest-neighbor (1N) and second-neighbor (2N) targets are colored in red.
    }
    \label{fig:graph}
\end{figure}
%
% \begin{figure}[tbph]
%     \centering
%     \includegraphics[width=.99\columnwidth ]{Figures/asymp_beta4.pdf}
%     \caption{
%     (a) Scaling behavior of the maximum $R(t_{opt})$ as a function of the size of the graph, for three different target positions: `C' central, `1N' nearest neighbor, `2N' second nearest neighbor. (b) Optimal time $t_{opt}$ as a function of $N$.
%     (c) Value of the target probabilities at the optimal time, for the quantum and classical search. The target detuning is set to $\beta/\gamma=4$.
%     }
%     \label{fig:asym}
% \end{figure}
\begin{figure}[tbph]
    \centering
    \includegraphics[width=.99\columnwidth ]{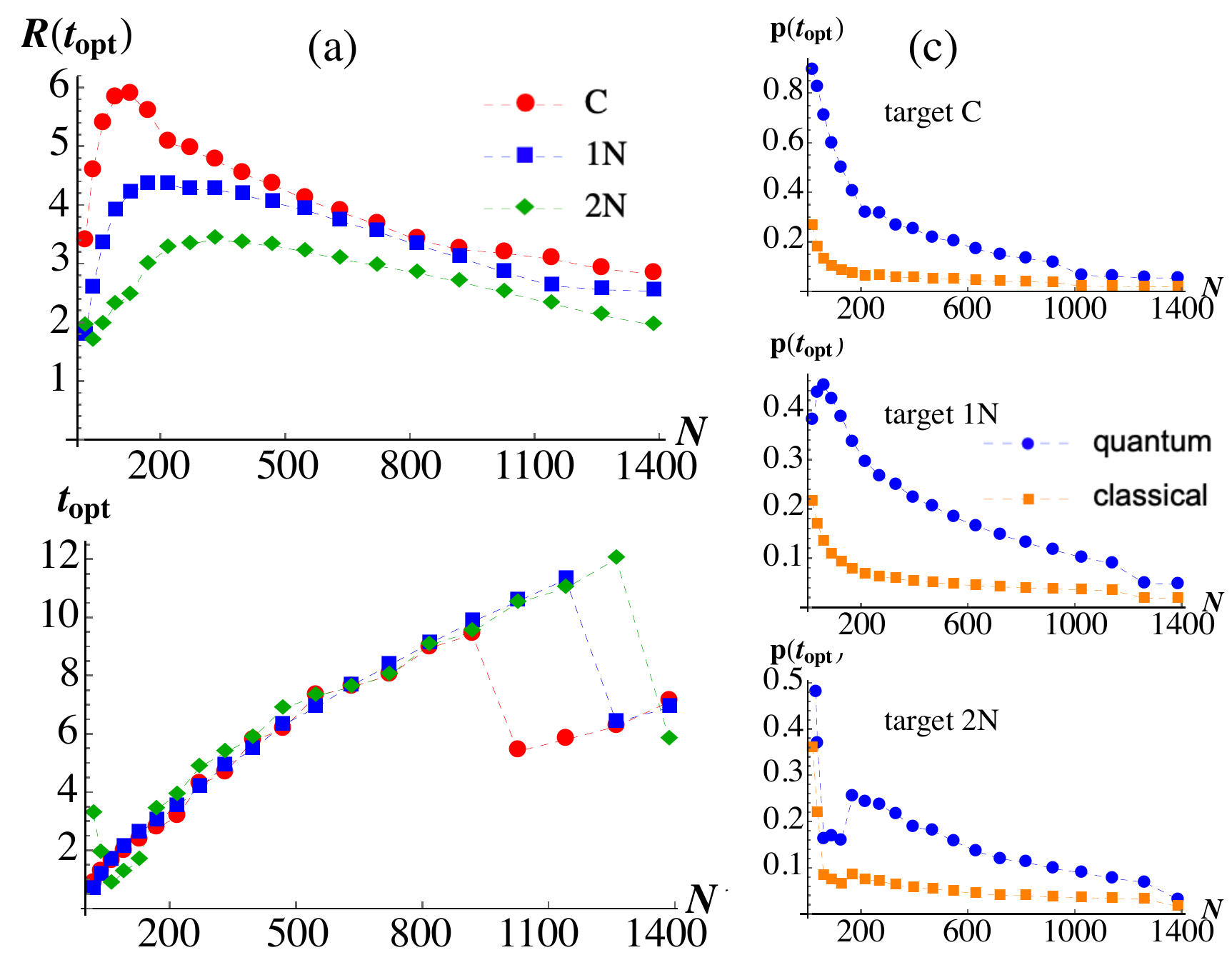}
    \caption{(a) Scaling behavior of the maximum $R(t_{opt})$ as a function of the size of the graph, for three different target positions: `C' central, `1N' nearest neighbor, `2N' second nearest neighbor. (b) Optimal time $t_{opt}$ as a function of $N$.
    (c) Value of the target probabilities at the optimal time, for the quantum and classical search. The target detuning is set to $\beta/\gamma=4$.
    }
    \label{fig:asym}
\end{figure}
\begin{figure*}[tbph]
    \centering
    \includegraphics[width=1.8\columnwidth ]{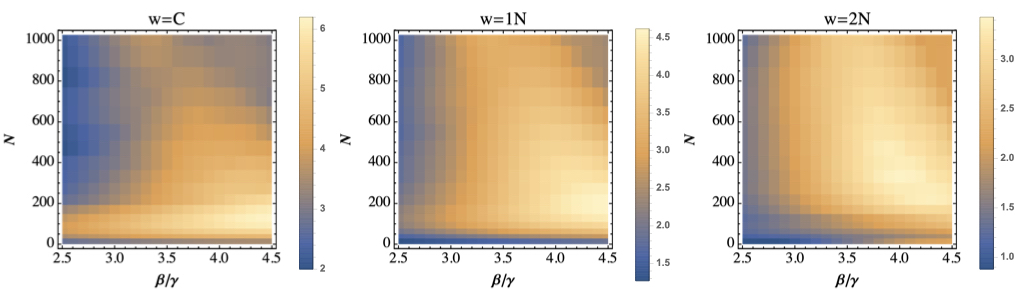}
    \caption{$R(t_\text{opt})$ as a function of the size $N$ of the graph and the parameter $\beta/\gamma$, for three different target positions: 'C' central, '1N' nearest neighbor, '2N' second nearest neighbor. 
    }
    \label{fig:variB}
\end{figure*}

In this section we show that the quantum advantage found in Section \ref{theosearch}
through the quantity $R(t)$ is still valid when the size of the graph is increased, i.e. the number of nodes $N$ gets larger.
{W}e consider finite-size triangular lattices, as sketched in Figure \ref{fig:graph}. 
The number of nodes is increased by adding  new external layers of vertices.
We focus on three positions for the target node: the central one (C),  nearest neighbour to the target (1N) and second-neighbor to the target (2N). 
Due to the symmetry of the considered topology, it is always possible to unequivocally identify the central target C and, in turn, its 1N and 2N nodes.
\newline
In Figure \ref{fig:asym}, we report the behaviour of the maximum value of the ratio $R(t_\text{opt})$ over time, i.e. 
\begin{equation}
   R(t_\text{opt})=\max_t R(t) .
\end{equation}
We call $t_\text{opt}$ the {\it optimal time}. As the number of nodes $N$ is increased, 
$R(t_\text{opt})$ remains larger than 1, indicating an advantage of the quantum search over the classical analogue (Fig. \ref{fig:asym}a); on the other hand, the optimal time increases with larger sizes of the lattice (Fig. \ref{fig:asym}b).
{
The sudden drop in $t_\text{opt}$ 
%for the C target 
is due to the presence of two competitive peaks in $R(t)$, independently on the target. As $N$  becomes large enough, the absolute maximum is reached  at shorter times.
%This behavior is not specific for the C node, but is found also for the other  targets as the size of the graph is increased.
}

In Fig.~\ref{fig:asym}c, we also show the behaviors of the quantum (blue) and classical (orange) target probabilities at the optimal time. Their values decreases with increasing $N$, but the probability of identifying the target with a quantum walker is always 
larger than with a classical walk. In Figure \ref{fig:variB} we display the optimal value of $R$ as a function of the target detuning $\beta/\gamma$ and of the increasing number of graph nodes $N$, for the three target {sites} considered.  The quantum  advantage, as quantified by $R(\gamma t)$,  is robust with respect to variations in $\beta/\gamma$ throughout  a wide region in the parameter-space, thus indicating that there is no need to precisely select a single optimal value for the detuning  in Eq. \eqref{hamQ}.

\section{Details of the fabricated waveguide arrays} \label{sec:appFab}

Waveguide arrays were inscribed in a 60~mm long fused-silica chip, according to the four sets of parameters (A, B, C, D) listed in Table~\ref{tab:ABCD}. As mentioned in Section~\ref{sec:spatialSearchWg} , three arrays were realized for each of these set of parameters: one without target (i.e. with all identical waveguides), one with the target on the site C, one with the target shifted on the site S. The waveguide separation $a$ and inscription speeds $v$ and $v_0$ were chosen to give the nominal coupling coefficient $\gamma$ and target detuning $\beta$ (where target is present) listed in the Table, according to preliminary calibration experiments. Note that the nominal ratio $\beta/\gamma$ is kept fixed to about 4.1.

The glass chip was later cut into two parts, with lengths of about 20~mm and 40~mm respectively. End facets were polished at optical quality (this further removed a few hundreds of microns). In this way, devices were obtained spanning eight different values of the normalized propagation coordinate (ranging from 0.77 to 2.32) as shown in Table~\ref{tab:afterCut}.  

\begin{table}[h!]%tbp]
\centering
\begin{tabular}{cccccc}
\toprule
\multirow{2}{*}{\textbf{Type}} & \multicolumn{3}{c}{\textbf{fabrication parameters}} & \multicolumn{2}{c}{\begin{tabular}{c}\textbf{nominal}\\\textbf{optical parameters}\end{tabular}}\\[11pt]
 & $a$~($\mu$m) & $v$~(mm/s) & $v_0$~(mm/s) & $\gamma$~(1/mm)& $\beta$~(1/mm)\\
\midrule
A & 23.40 & 3.69 & 2.00 & 0.060 & 0.25\\
B & 24.37 & 3.32 & 2.00 & 0.053 & 0.22\\
C & 25.30 & 3.03 & 2.00 & 0.047 & 0.19\\
D & 26.56 & 2.78 & 2.00 & 0.040 & 0.17\\
\bottomrule
\end{tabular}
\caption{\label{tab:ABCD} List of the geometrical and optical parameters of the fabricated arrays. $a$ is the interwaveguide distance, $v$ is the inscription speed of all waveguides except the target one, $v_0$ is the inscription speed of the target waveguide. $\gamma$ is the nominal coupling coefficient and $\beta$ is the nominal target detuning, which are predicted from preliminary calibration experiments, given the values of $a$, $v$ and $v_0$.}
\end{table}

\begin{table}[h!]%tbp]
\centering
\begin{tabular}{ccc}
\toprule
\multirow{2}{*}{\textbf{Type}} & \multicolumn{2}{c}{\textbf{nominal}~$\gamma t$}\\
 & $t$~=~19.3~mm & $t$~=~38.6~mm\\
 \midrule
 A & 1.16 & 2.32\\
 B & 1.02 & 2.05\\
 C & 0.91 & 1.81\\
 D & 0.77 & 1.54\\
  \bottomrule
\end{tabular}
\caption{\label{tab:afterCut}Values of the normalized propagation coordinate corresponding to the eight types of arrays available after cutting and polishing the laser-inscribed glass sample. For each of these eight types, three arrays are present, with the different target possibilities as in Fig.~\ref{fig:geometry}.}
\end{table}

\pagebreak
\section{Alignment of the waveguide array with the coherent illumination beam} \label{sec:appTilt}
In order to perform reliable measurements, the waveguide arrays must be accurately aligned  with respect to the laser beam, especially in the tilt degrees of freedom. To this purpose, the glass chip was mounted on a multi-axis micrometric manipulator (three translation and two rotation axes) and the alignment was operated according to the following procedure. 

As a first thing, one uniform array is selected and centered with respect to the input laser beam. A rough adjustment of the transverse position and inclination of the glass chip is conducted without any collection objective, by overlapping the far-field interference pattern produced by the array output with the far-field spot of the uncoupled light, while maximizing the intensity of the former. 

In fact, part of the light focused by the $f$~=~40~cm lens is not effectively coupled into the waveguides, but keeps propagating and diverging, first within the glass and then out of it. The light coupled into the waveguides and then exiting at the array output, instead, produces a complex interference pattern in the far-field. On the one side, by micrometrically adjusting the sample transverse position, one can maximize the intensity of such interference pattern and thus optimize the overlap of the impinging wavefront with the optical modes of the array. On the other side, action on the tilt degrees of freedom has the effect of translating the far-field position of the interference pattern. If the latter is overlapped to the spot of the uncoupled light, this means that the uncoupled light has the same average wavevector as the light propagating in the waveguides (and then out of them). Given the symmetry of the array, this also imply that the desired input condition has been reached, with all modes excited with the same phase.

\begin{figure}[tbp]
\includegraphics[scale=1]{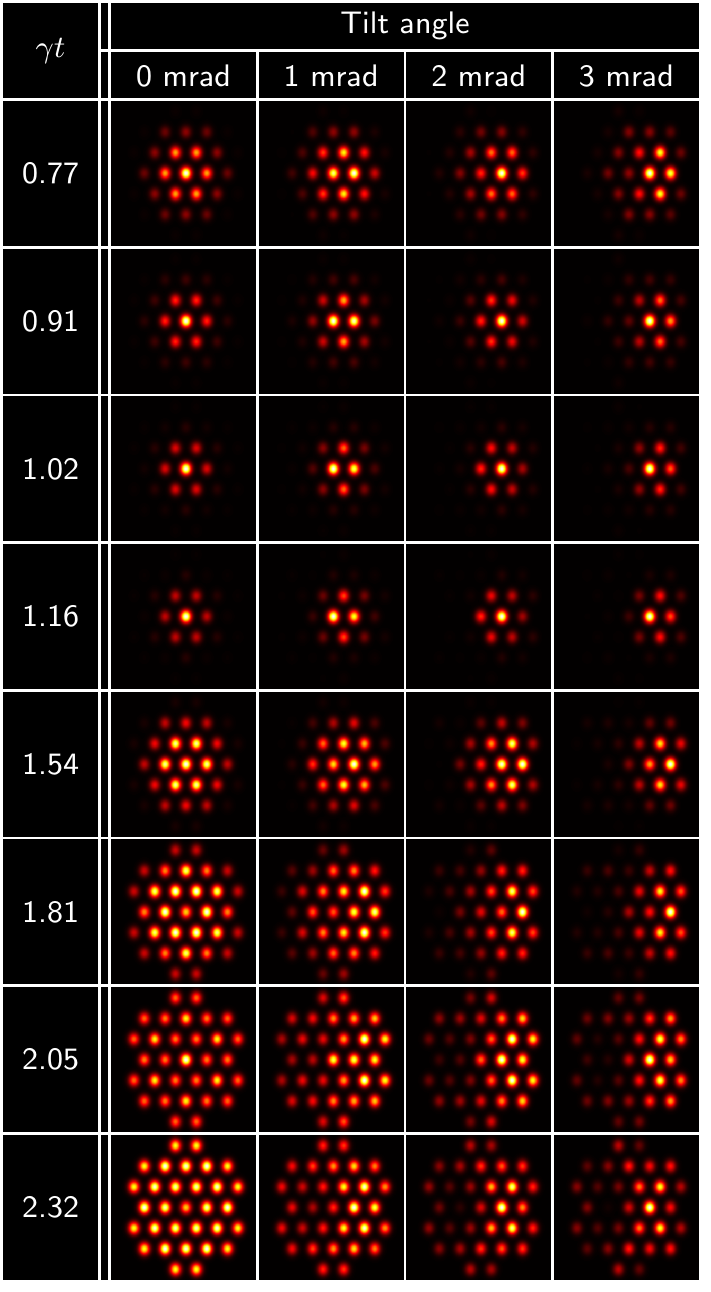}
\caption{\label{fig:tilting} Numerical simulation of the propagation in waveguide arrays with the geometry as in Fig.~\ref{fig:geometry}, with an interwaveguide spacing $a=25\;\mu\mathrm{m}$ and no target site implemented (all waveguides are equal), excited at the input with a plane wave with different tilt angles (wavelength 633~nm).  The intensity scale of each picture is normalized to the maximum value of that picture. One notes that even a small tilt angle can alter significantly the light distribution.}
\end{figure}

The alignment is completed by placing the collection objective in front of the array output, and by imaging the output facet onto a CCD camera. The sample tilt is now finely tuned in order to achieve the most symmetric output distribution {(see also Fig.~\ref{fig:tilting})}. 

Since the arrays are inscribed parallel one to the other, we could observe that, once the tilts are optimized for one device, the alignment can be safely maintained also for the neighbouring ones, which can be coupled by just a rigid translation of the glass sample. Thus, the uniform array serves for alignment purposes, and allows for reliable measurements on the neighbouring two, which are interesting for the spatial search experiment.

\begin{figure}[tbp]
\includegraphics[scale=1]{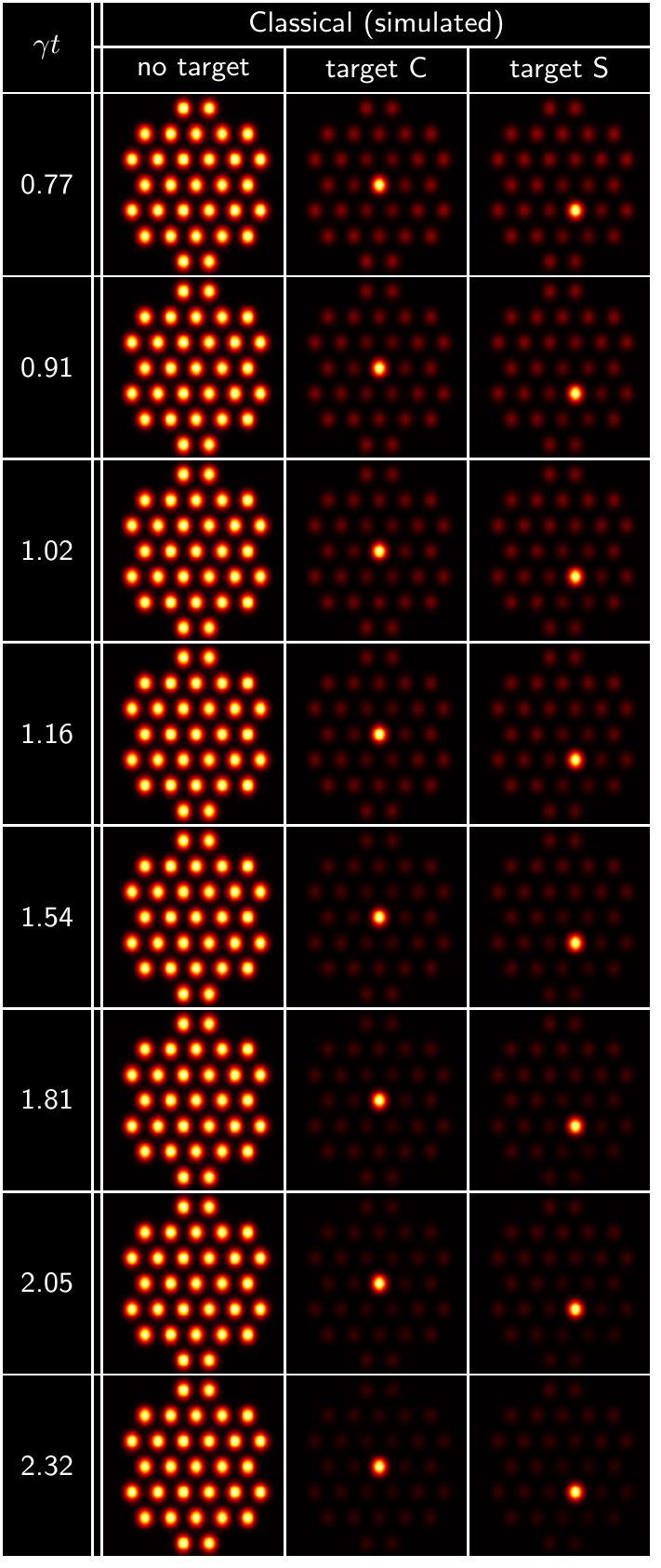}
\caption{\label{fig:classical} Numerical simulation of a classical random walk on a 31-sites graph, as described in Section II. We consider the case of the absence of target, target on the central site C and target on a site shifted from center S as in Fig.~\ref{fig:geometry}. At the initial time $t=0$ the walker populates with identical probabilities all sites. We have chosen to represent the probability distribution of the walker at each $\gamma t$ in a graphical way that matches the optical intensity distributions in Fig.~\ref{fig:exp}, to facilitate comparison. The intensity scale of each picture is normalized to the maximum value of that picture.}
\end{figure}

%\clearpage


\begin{thebibliography}{41}%
\makeatletter
\providecommand \@ifxundefined [1]{%
 \@ifx{#1\undefined}
}%
\providecommand \@ifnum [1]{%
 \ifnum #1\expandafter \@firstoftwo
 \else \expandafter \@secondoftwo
 \fi
}%
\providecommand \@ifx [1]{%
 \ifx #1\expandafter \@firstoftwo
 \else \expandafter \@secondoftwo
 \fi
}%
\providecommand \natexlab [1]{#1}%
\providecommand \enquote  [1]{``#1''}%
\providecommand \bibnamefont  [1]{#1}%
\providecommand \bibfnamefont [1]{#1}%
\providecommand \citenamefont [1]{#1}%
\providecommand \href@noop [0]{\@secondoftwo}%
\providecommand \href [0]{\begingroup \@sanitize@url \@href}%
\providecommand \@href[1]{\@@startlink{#1}\@@href}%
\providecommand \@@href[1]{\endgroup#1\@@endlink}%
\providecommand \@sanitize@url [0]{\catcode `\\12\catcode `\$12\catcode
  `\&12\catcode `\#12\catcode `\^12\catcode `\_12\catcode `\%12\relax}%
\providecommand \@@startlink[1]{}%
\providecommand \@@endlink[0]{}%
\providecommand \url  [0]{\begingroup\@sanitize@url \@url }%
\providecommand \@url [1]{\endgroup\@href {#1}{\urlprefix }}%
\providecommand \urlprefix  [0]{URL }%
\providecommand \Eprint [0]{\href }%
\providecommand \doibase [0]{http://dx.doi.org/}%
\providecommand \selectlanguage [0]{\@gobble}%
\providecommand \bibinfo  [0]{\@secondoftwo}%
\providecommand \bibfield  [0]{\@secondoftwo}%
\providecommand \translation [1]{[#1]}%
\providecommand \BibitemOpen [0]{}%
\providecommand \bibitemStop [0]{}%
\providecommand \bibitemNoStop [0]{.\EOS\space}%
\providecommand \EOS [0]{\spacefactor3000\relax}%
\providecommand \BibitemShut  [1]{\csname bibitem#1\endcsname}%
\let\auto@bib@innerbib\@empty
%</preamble>
\bibitem [{\citenamefont {Chen}\ and\ \citenamefont {Zhan}(2008)}]{chen08}%
  \BibitemOpen
  \bibfield  {author} {\bibinfo {author} {\bibfnamefont {H.}~\bibnamefont
  {Chen}}\ and\ \bibinfo {author} {\bibfnamefont {F.}~\bibnamefont {Zhan}},\
  }\bibfield  {title} {\enquote {\bibinfo {title} {The expected hitting times
  for finite {M}arkov chains},}\ }\href {\doibase
  https://doi.org/10.1016/j.laa.2008.01.003} {\bibfield  {journal} {\bibinfo
  {journal} {Lin. Alg. Appl. (2008) 2730–2749%Available online at
  www.sciencedirect.comwww.elsevier.com/locate/laa
  }\ }\textbf {\bibinfo
  {volume} {428}},\ \bibinfo {pages} {2730--2749} (\bibinfo {year}
  {2008})}\BibitemShut {NoStop}%
\bibitem [{\citenamefont {Kempe}(2003{\natexlab{a}})}]{kempe003}%
  \BibitemOpen
  \bibfield  {author} {\bibinfo {author} {\bibfnamefont {J.}~\bibnamefont
  {Kempe}},\ }\bibfield  {title} {\enquote {\bibinfo {title} {Quantum random
  walks: An introductory overview},}\ }\href {\doibase
  10.1080/00107151031000110776} {\bibfield  {journal} {\bibinfo  {journal}
  {Contemp. Phys.}\ }\textbf {\bibinfo {volume} {44}},\ \bibinfo {pages}
  {307--327} (\bibinfo {year} {2003}{\natexlab{a}})}\BibitemShut {NoStop}%
\bibitem [{\citenamefont {Venegas-Andraca}(2012)}]{Venegas12}%
  \BibitemOpen
  \bibfield  {author} {\bibinfo {author} {\bibfnamefont {S.~E.}\ \bibnamefont
  {Venegas-Andraca}},\ }\bibfield  {title} {\enquote {\bibinfo {title} {Quantum
  walks: a comprehensive review},}\ }\href {\doibase 10.1007/s11128-012-0432-5}
  {\bibfield  {journal} {\bibinfo  {journal} {Quantum Inf. Proc.}\ }\textbf
  {\bibinfo {volume} {11}},\ \bibinfo {pages} {1015--1106} (\bibinfo {year}
  {2012})}\BibitemShut {NoStop}%
\bibitem [{\citenamefont {Childs}\ \emph {et~al.}(2002)\citenamefont {Childs},
  \citenamefont {Farhi},\ and\ \citenamefont {Gutmann}}]{childs02}%
  \BibitemOpen
  \bibfield  {author} {\bibinfo {author} {\bibfnamefont {A.~M.}\ \bibnamefont
  {Childs}}, \bibinfo {author} {\bibfnamefont {E.}~\bibnamefont {Farhi}}, \
  and\ \bibinfo {author} {\bibfnamefont {S.}~\bibnamefont {Gutmann}},\
  }\bibfield  {title} {\enquote {\bibinfo {title} {An example of the difference
  between quantum and classical random walks},}\ }\href {\doibase
  10.1023/A:1019609420309} {\bibfield  {journal} {\bibinfo  {journal} {Quantum
  Inf. Proc.}\ }\textbf {\bibinfo {volume} {1}},\ \bibinfo {pages}
  {35--43} (\bibinfo {year} {2002})}\BibitemShut {NoStop}%
\bibitem [{\citenamefont {Gualtieri}\ \emph {et~al.}(2020)\citenamefont
  {Gualtieri}, \citenamefont {Benedetti},\ and\ \citenamefont
  {Paris}}]{gualtieri20}%
  \BibitemOpen
  \bibfield  {author} {\bibinfo {author} {\bibfnamefont {V.}~\bibnamefont
  {Gualtieri}}, \bibinfo {author} {\bibfnamefont {C.}~\bibnamefont
  {Benedetti}}, \ and\ \bibinfo {author} {\bibfnamefont {M.~G.~A.}\
  \bibnamefont {Paris}},\ }\bibfield  {title} {\enquote {\bibinfo {title}
  {Quantum-classical dynamical distance and quantumness of quantum walks},}\
  }\href {\doibase 10.1103/PhysRevA.102.012201} {\bibfield  {journal} {\bibinfo
   {journal} {Phys. Rev. A}\ }\textbf {\bibinfo {volume} {102}},\ \bibinfo
  {pages} {012201} (\bibinfo {year} {2020})}\BibitemShut {NoStop}%
\bibitem [{\citenamefont {Childs}\ \emph {et~al.}(2003)\citenamefont {Childs},
  \citenamefont {Cleve}, \citenamefont {Deotto}, \citenamefont {Farhi},
  \citenamefont {Gutmann},\ and\ \citenamefont {Spielman}}]{childs03}%
  \BibitemOpen
  \bibfield  {author} {\bibinfo {author} {\bibfnamefont {A.~M.}\ \bibnamefont
  {Childs}}, \bibinfo {author} {\bibfnamefont {R.}~\bibnamefont {Cleve}},
  \bibinfo {author} {\bibfnamefont {E.}~\bibnamefont {Deotto}}, \bibinfo
  {author} {\bibfnamefont {E.}~\bibnamefont {Farhi}}, \bibinfo {author}
  {\bibfnamefont {S.}~\bibnamefont {Gutmann}}, \ and\ \bibinfo {author}
  {\bibfnamefont {D.~A.}\ \bibnamefont {Spielman}},\ }\bibfield  {title}
  {\enquote {\bibinfo {title} {Exponential algorithmic speedup by a quantum
  walk},}\ }in\ \href {\doibase 10.1145/780542.780552} {\emph {\bibinfo
  {booktitle} {Proceedings of the Thirty-Fifth Annual ACM Symposium on Theory
  of Computing}}},\ \bibinfo {series and number} {STOC '03}\ (\bibinfo
  {publisher} {Association for Computing Machinery},\ \bibinfo {address} {New
  York, NY, USA},\ \bibinfo {year} {2003})\ p.\ \bibinfo {pages}
  {59–68}\BibitemShut {NoStop}%
\bibitem [{\citenamefont {Farhi}\ and\ \citenamefont
  {Gutmann}(1998)}]{farhi98}%
  \BibitemOpen
  \bibfield  {author} {\bibinfo {author} {\bibfnamefont {E.}~\bibnamefont
  {Farhi}}\ and\ \bibinfo {author} {\bibfnamefont {S.}~\bibnamefont
  {Gutmann}},\ }\bibfield  {title} {\enquote {\bibinfo {title} {Quantum
  computation and decision trees},}\ }\href {\doibase 10.1103/PhysRevA.58.915}
  {\bibfield  {journal} {\bibinfo  {journal} {Phys. Rev. A}\ }\textbf {\bibinfo
  {volume} {58}},\ \bibinfo {pages} {915--928} (\bibinfo {year}
  {1998})}\BibitemShut {NoStop}%
\bibitem [{\citenamefont {Childs}\ and\ \citenamefont
  {Goldstone}(2004)}]{childs2004}%
  \BibitemOpen
  \bibfield  {author} {\bibinfo {author} {\bibfnamefont {A.~M.}\ \bibnamefont
  {Childs}}\ and\ \bibinfo {author} {\bibfnamefont {J.}~\bibnamefont
  {Goldstone}},\ }\bibfield  {title} {\enquote {\bibinfo {title} {Spatial
  search by quantum walk},}\ }\href {\doibase 10.1103/PhysRevA.70.022314}
  {\bibfield  {journal} {\bibinfo  {journal} {Phys. Rev. A}\ }\textbf {\bibinfo
  {volume} {70}},\ \bibinfo {pages} {022314} (\bibinfo {year}
  {2004})}\BibitemShut {NoStop}%
\bibitem [{\citenamefont {Childs}(2009)}]{childs09}%
  \BibitemOpen
  \bibfield  {author} {\bibinfo {author} {\bibfnamefont {A.~M.}\ \bibnamefont
  {Childs}},\ }\bibfield  {title} {\enquote {\bibinfo {title} {Universal
  computation by quantum walk},}\ }\href {\doibase
  10.1103/PhysRevLett.102.180501} {\bibfield  {journal} {\bibinfo  {journal}
  {Phys. Rev. Lett.}\ }\textbf {\bibinfo {volume} {102}},\ \bibinfo {pages}
  {180501} (\bibinfo {year} {2009})}\BibitemShut {NoStop}%
\bibitem [{\citenamefont {de~Falco}\ and\ \citenamefont
  {Tamascelli}(2013)}]{tama13}%
  \BibitemOpen
  \bibfield  {author} {\bibinfo {author} {\bibfnamefont {D.}~\bibnamefont
  {de~Falco}}\ and\ \bibinfo {author} {\bibfnamefont {D.}~\bibnamefont
  {Tamascelli}},\ }\bibfield  {title} {\enquote {\bibinfo {title}
  {Noise-assisted quantum transport and computation},}\ }\href {\doibase 	10.1088/1751-8113/46/22/225301}
  {\bibfield  {journal} {\bibinfo  {journal} {J. Phys. A: Math. Theor.}\
  }\textbf {\bibinfo {volume} {46}},\ \bibinfo {pages} {225301} (\bibinfo
  {year} {2013})}\BibitemShut {NoStop}%
\bibitem [{\citenamefont {Tamascelli}\ and\ \citenamefont
  {Zanetti}(2014)}]{tama14}%
  \BibitemOpen
  \bibfield  {author} {\bibinfo {author} {\bibfnamefont {D.}~\bibnamefont
  {Tamascelli}}\ and\ \bibinfo {author} {\bibfnamefont {L.}~\bibnamefont
  {Zanetti}},\ }\bibfield  {title} {\enquote {\bibinfo {title} {A
  quantum-walk-inspired adiabatic algorithm for solving graph isomorphism
  problems},}\ }\href {\doibase 10.1088/1751-8113/47/32/325302} {\bibfield
  {journal} {\bibinfo  {journal} {J. Phys. A: Math.
  Theor.}\ }\textbf {\bibinfo {volume} {47}},\ \bibinfo {pages} {325302}
  (\bibinfo {year} {2014})}\BibitemShut {NoStop}%
\bibitem [{\citenamefont {Tamascelli}\ \emph {et~al.}(2016)\citenamefont
  {Tamascelli}, \citenamefont {Olivares}, \citenamefont {Rossotti},
  \citenamefont {Osellame},\ and\ \citenamefont {Paris}}]{tama16}%
  \BibitemOpen
  \bibfield  {author} {\bibinfo {author} {\bibfnamefont {D.}~\bibnamefont
  {Tamascelli}}, \bibinfo {author} {\bibfnamefont {S.}~\bibnamefont
  {Olivares}}, \bibinfo {author} {\bibfnamefont {S.}~\bibnamefont {Rossotti}},
  \bibinfo {author} {\bibfnamefont {R.}~\bibnamefont {Osellame}}, \ and\
  \bibinfo {author} {\bibfnamefont {M.~G.~A.}\ \bibnamefont {Paris}},\
  }\bibfield  {title} {\enquote {\bibinfo {title} {Quantum state transfer via
  {Bloch} oscillations},}\ }\href {https://doi.org/10.1038/srep26054}
  {\bibfield  {journal} {\bibinfo  {journal} {Sci. Rep.}\ }\textbf {\bibinfo
  {volume} {6}},\ \bibinfo {pages} {26054} (\bibinfo {year}
  {2016})}\BibitemShut {NoStop}%
\bibitem [{\citenamefont {Chakraborty}\ \emph {et~al.}(2020)\citenamefont
  {Chakraborty}, \citenamefont {Novo},\ and\ \citenamefont
  {Roland}}]{chakra20}%
  \BibitemOpen
  \bibfield  {author} {\bibinfo {author} {\bibfnamefont {S.}~\bibnamefont
  {Chakraborty}}, \bibinfo {author} {\bibfnamefont {L.}~\bibnamefont {Novo}}, \
  and\ \bibinfo {author} {\bibfnamefont {J.}~\bibnamefont {Roland}},\
  }\bibfield  {title} {\enquote {\bibinfo {title} {Optimality of spatial search
  via continuous-time quantum walks},}\ }\href {\doibase
  10.1103/PhysRevA.102.032214} {\bibfield  {journal} {\bibinfo  {journal}
  {Phys. Rev. A}\ }\textbf {\bibinfo {volume} {102}},\ \bibinfo {pages}
  {032214} (\bibinfo {year} {2020})}\BibitemShut {NoStop}%
\bibitem [{\citenamefont {Paris}\ \emph {et~al.}(2021)\citenamefont {Paris},
  \citenamefont {Benedetti},\ and\ \citenamefont {Olivares}}]{sym13010096}%
  \BibitemOpen
  \bibfield  {author} {\bibinfo {author} {\bibfnamefont {M.~G.~A.}\
  \bibnamefont {Paris}}, \bibinfo {author} {\bibfnamefont {C.}~\bibnamefont
  {Benedetti}}, \ and\ \bibinfo {author} {\bibfnamefont {S.}~\bibnamefont
  {Olivares}},\ }\bibfield  {title} {\enquote {\bibinfo {title} {Improving
  quantum search on simple graphs by pretty good structured oracles},}\ }\href
  {\doibase 10.3390/sym13010096} {\bibfield  {journal} {\bibinfo  {journal}
  {Symmetry}\ }\textbf {\bibinfo {volume} {13}} (\bibinfo {year} {2021}) %,\  10.3390/sym13010096
  }\BibitemShut {NoStop}%
\bibitem [{\citenamefont {Aharonov}\ \emph {et~al.}(1993)\citenamefont
  {Aharonov}, \citenamefont {Davidovich},\ and\ \citenamefont
  {Zagury}}]{aharonov93}%
  \BibitemOpen
  \bibfield  {author} {\bibinfo {author} {\bibfnamefont {Y.}~\bibnamefont
  {Aharonov}}, \bibinfo {author} {\bibfnamefont {L.}~\bibnamefont
  {Davidovich}}, \ and\ \bibinfo {author} {\bibfnamefont {N.}~\bibnamefont
  {Zagury}},\ }\bibfield  {title} {\enquote {\bibinfo {title} {Quantum random
  walks},}\ }\href {\doibase 10.1103/PhysRevA.48.1687} {\bibfield  {journal}
  {\bibinfo  {journal} {Phys. Rev. A}\ }\textbf {\bibinfo {volume} {48}},\
  \bibinfo {pages} {1687--1690} (\bibinfo {year} {1993})}\BibitemShut {NoStop}%
\bibitem [{\citenamefont {Kempe}(2003{\natexlab{b}})}]{kempe03}%
  \BibitemOpen
  \bibfield  {author} {\bibinfo {author} {\bibfnamefont {J.}~\bibnamefont
  {Kempe}},\ }\bibfield  {title} {\enquote {\bibinfo {title} {Discrete quantum
  walks hit exponentially faster},}\ }in\ \href@noop {} {\emph {\bibinfo
  {booktitle} {Approximation, Randomization, and Combinatorial Optimization..
  Algorithms and Techniques}}},\ \bibinfo {editor} {edited by\ \bibinfo
  {editor} {\bibfnamefont {S.}~\bibnamefont {Arora}}, \bibinfo {editor}
  {\bibfnamefont {K.}~\bibnamefont {Jansen}}, \bibinfo {editor} {\bibfnamefont
  {J.~D.~P.}\ \bibnamefont {Rolim}}, \ and\ \bibinfo {editor} {\bibfnamefont
  {A.}~\bibnamefont {Sahai}}}\ (\bibinfo  {publisher} {Springer Berlin
  Heidelberg},\ \bibinfo {address} {Berlin, Heidelberg},\ \bibinfo {year}
  {2003})\ pp.\ \bibinfo {pages} {354--369}\BibitemShut {NoStop}%
\bibitem [{\citenamefont {Shenvi}\ \emph {et~al.}(2003)\citenamefont {Shenvi},
  \citenamefont {Kempe},\ and\ \citenamefont {Whaley}}]{shenvi03}%
  \BibitemOpen
  \bibfield  {author} {\bibinfo {author} {\bibfnamefont {N.}~\bibnamefont
  {Shenvi}}, \bibinfo {author} {\bibfnamefont {J.}~\bibnamefont {Kempe}}, \
  and\ \bibinfo {author} {\bibfnamefont {K.~B.}\ \bibnamefont {Whaley}},\
  }\bibfield  {title} {\enquote {\bibinfo {title} {Quantum random-walk search
  algorithm},}\ }\href {\doibase 10.1103/PhysRevA.67.052307} {\bibfield
  {journal} {\bibinfo  {journal} {Phys. Rev. A}\ }\textbf {\bibinfo {volume}
  {67}},\ \bibinfo {pages} {052307} (\bibinfo {year} {2003})}\BibitemShut
  {NoStop}%
\bibitem [{\citenamefont {Lovett}\ \emph {et~al.}(2010)\citenamefont {Lovett},
  \citenamefont {Cooper}, \citenamefont {Everitt}, \citenamefont {Trevers},\
  and\ \citenamefont {Kendon}}]{lovett10}%
  \BibitemOpen
  \bibfield  {author} {\bibinfo {author} {\bibfnamefont {N.~B.}\ \bibnamefont
  {Lovett}}, \bibinfo {author} {\bibfnamefont {S.}~\bibnamefont {Cooper}},
  \bibinfo {author} {\bibfnamefont {M.}~\bibnamefont {Everitt}}, \bibinfo
  {author} {\bibfnamefont {M.}~\bibnamefont {Trevers}}, \ and\ \bibinfo
  {author} {\bibfnamefont {V.}~\bibnamefont {Kendon}},\ }\bibfield  {title}
  {\enquote {\bibinfo {title} {Universal quantum computation using the
  discrete-time quantum walk},}\ }\href {\doibase 10.1103/PhysRevA.81.042330}
  {\bibfield  {journal} {\bibinfo  {journal} {Phys. Rev. A}\ }\textbf {\bibinfo
  {volume} {81}},\ \bibinfo {pages} {042330} (\bibinfo {year}
  {2010})}\BibitemShut {NoStop}%
\bibitem [{\citenamefont {Koch}\ and\ \citenamefont {Hillery}(2018)}]{koch18}%
  \BibitemOpen
  \bibfield  {author} {\bibinfo {author} {\bibfnamefont {D.}~\bibnamefont
  {Koch}}\ and\ \bibinfo {author} {\bibfnamefont {M.}~\bibnamefont {Hillery}},\
  }\bibfield  {title} {\enquote {\bibinfo {title} {Finding paths in tree graphs
  with a quantum walk},}\ }\href {\doibase 10.1103/PhysRevA.97.012308}
  {\bibfield  {journal} {\bibinfo  {journal} {Phys. Rev. A}\ }\textbf {\bibinfo
  {volume} {97}},\ \bibinfo {pages} {012308} (\bibinfo {year}
  {2018})}\BibitemShut {NoStop}%
\bibitem [{\citenamefont {Childs}\ and\ \citenamefont {Ge}(2014)}]{childs14}%
  \BibitemOpen
  \bibfield  {author} {\bibinfo {author} {\bibfnamefont {A.~M.}\ \bibnamefont
  {Childs}}\ and\ \bibinfo {author} {\bibfnamefont {Y.}~\bibnamefont {Ge}},\
  }\bibfield  {title} {\enquote {\bibinfo {title} {Spatial search by
  continuous-time quantum walks on crystal lattices},}\ }\href {\doibase
  10.1103/PhysRevA.89.052337} {\bibfield  {journal} {\bibinfo  {journal} {Phys.
  Rev. A}\ }\textbf {\bibinfo {volume} {89}},\ \bibinfo {pages} {052337}
  (\bibinfo {year} {2014})}\BibitemShut {NoStop}%
\bibitem [{\citenamefont {Meyer}\ and\ \citenamefont {Wong}(2015)}]{meyer15}%
  \BibitemOpen
  \bibfield  {author} {\bibinfo {author} {\bibfnamefont {D.~A.}\ \bibnamefont
  {Meyer}}\ and\ \bibinfo {author} {\bibfnamefont {T.~G.}\ \bibnamefont
  {Wong}},\ }\bibfield  {title} {\enquote {\bibinfo {title} {Connectivity is a
  poor indicator of fast quantum search},}\ }\href {\doibase
  10.1103/PhysRevLett.114.110503} {\bibfield  {journal} {\bibinfo  {journal}
  {Phys. Rev. Lett.}\ }\textbf {\bibinfo {volume} {114}},\ \bibinfo {pages}
  {110503} (\bibinfo {year} {2015})}\BibitemShut {NoStop}%
\bibitem [{\citenamefont {Philipp}\ \emph {et~al.}(2016)\citenamefont
  {Philipp}, \citenamefont {Tarrataca},\ and\ \citenamefont
  {Boettcher}}]{philipp16}%
  \BibitemOpen
  \bibfield  {author} {\bibinfo {author} {\bibfnamefont {P.}~\bibnamefont
  {Philipp}}, \bibinfo {author} {\bibfnamefont {L.}~\bibnamefont {Tarrataca}},
  \ and\ \bibinfo {author} {\bibfnamefont {S}~\bibnamefont {Boettcher}},\
  }\bibfield  {title} {\enquote {\bibinfo {title} {Continuous-time quantum
  search on balanced trees},}\ }\href {\doibase 10.1103/PhysRevA.93.032305}
  {\bibfield  {journal} {\bibinfo  {journal} {Phys. Rev. A}\ }\textbf {\bibinfo
  {volume} {93}},\ \bibinfo {pages} {032305} (\bibinfo {year}
  {2016})}\BibitemShut {NoStop}%
\bibitem [{\citenamefont {Chakraborty}\ \emph {et~al.}(2016)\citenamefont
  {Chakraborty}, \citenamefont {Novo}, \citenamefont {Ambainis},\ and\
  \citenamefont {Omar}}]{yasser16}%
  \BibitemOpen
  \bibfield  {author} {\bibinfo {author} {\bibfnamefont {S.}~\bibnamefont
  {Chakraborty}}, \bibinfo {author} {\bibfnamefont {L.}~\bibnamefont {Novo}},
  \bibinfo {author} {\bibfnamefont {A.}~\bibnamefont {Ambainis}}, \ and\
  \bibinfo {author} {\bibfnamefont {Y.}~\bibnamefont {Omar}},\ }\bibfield
  {title} {\enquote {\bibinfo {title} {Spatial search by quantum walk is
  optimal for almost all graphs},}\ }\href {\doibase
  10.1103/PhysRevLett.116.100501} {\bibfield  {journal} {\bibinfo  {journal}
  {Phys. Rev. Lett.}\ }\textbf {\bibinfo {volume} {116}},\ \bibinfo {pages}
  {100501} (\bibinfo {year} {2016})}\BibitemShut {NoStop}%
\bibitem [{\citenamefont {Osada}\ \emph {et~al.}(2018)\citenamefont {Osada},
  \citenamefont {Sanaka}, \citenamefont {Munro},\ and\ \citenamefont
  {Nemoto}}]{osada18}%
  \BibitemOpen
  \bibfield  {author} {\bibinfo {author} {\bibfnamefont {T.}~\bibnamefont
  {Osada}}, \bibinfo {author} {\bibfnamefont {K.}~\bibnamefont {Sanaka}},
  \bibinfo {author} {\bibfnamefont {W.~J.}\ \bibnamefont {Munro}}, \ and\
  \bibinfo {author} {\bibfnamefont {Kae}\ \bibnamefont {Nemoto}},\ }\bibfield
  {title} {\enquote {\bibinfo {title} {Spatial search on a two-dimensional
  lattice with long-range interactions},}\ }\href {\doibase
  10.1103/PhysRevA.97.062319} {\bibfield  {journal} {\bibinfo  {journal} {Phys.
  Rev. A}\ }\textbf {\bibinfo {volume} {97}},\ \bibinfo {pages} {062319}
  (\bibinfo {year} {2018})}\BibitemShut {NoStop}%
\bibitem [{\citenamefont {Glos}\ and\ \citenamefont {Januszek}(2019)}]{glos19}%
  \BibitemOpen
  \bibfield  {author} {\bibinfo {author} {\bibfnamefont {A.}~\bibnamefont
  {Glos}}\ and\ \bibinfo {author} {\bibfnamefont {T.}~\bibnamefont
  {Januszek}},\ }\bibfield  {title} {\enquote {\bibinfo {title} {Impact of
  global and local interaction on quantum spatial search on chimera graph},}\
  }\href {\doibase 10.1142/S0219749919500400} {\bibfield  {journal} {\bibinfo
  {journal} {Int. J. Quantum Inf.}\ }\textbf {\bibinfo {volume} {17}},\
  \bibinfo {pages} {1950040} (\bibinfo {year} {2019})}\BibitemShut {NoStop}%
\bibitem [{\citenamefont {Osada}\ \emph {et~al.}(2020)\citenamefont {Osada},
  \citenamefont {Coutinho}, \citenamefont {Omar}, \citenamefont {Sanaka},
  \citenamefont {Munro},\ and\ \citenamefont {Nemoto}}]{osada20}%
  \BibitemOpen
  \bibfield  {author} {\bibinfo {author} {\bibfnamefont {T.}~\bibnamefont
  {Osada}}, \bibinfo {author} {\bibfnamefont {B.}~\bibnamefont {Coutinho}},
  \bibinfo {author} {\bibfnamefont {Y.}~\bibnamefont {Omar}}, \bibinfo {author}
  {\bibfnamefont {K.}~\bibnamefont {Sanaka}}, \bibinfo {author} {\bibfnamefont
  {W.~J.}\ \bibnamefont {Munro}}, \ and\ \bibinfo {author} {\bibfnamefont
  {K.}~\bibnamefont {Nemoto}},\ }\bibfield  {title} {\enquote {\bibinfo {title}
  {Continuous-time quantum-walk spatial search on the bollob\'as scale-free
  network},}\ }\href {\doibase 10.1103/PhysRevA.101.022310} {\bibfield
  {journal} {\bibinfo  {journal} {Phys. Rev. A}\ }\textbf {\bibinfo {volume}
  {101}},\ \bibinfo {pages} {022310} (\bibinfo {year} {2020})}\BibitemShut
  {NoStop}%
\bibitem [{\citenamefont {Portugal}(2018)}]{portugal18}%
  \BibitemOpen
  \bibfield  {author} {\bibinfo {author} {\bibfnamefont {R.}~\bibnamefont
  {Portugal}},\ }\href {\doibase 10.1007/978-3-319-97813-0} {\emph {\bibinfo
  {title} {Quantum Walks and Search Algorithms}}}\ (\bibinfo  {publisher}
  {Springer International Publishing},\ \bibinfo {year} {2018})\BibitemShut
  {NoStop}%
\bibitem [{\citenamefont {Foulger}\ \emph {et~al.}(2014)\citenamefont
  {Foulger}, \citenamefont {Gnutzmann},\ and\ \citenamefont
  {Tanner}}]{foulger14}%
  \BibitemOpen
  \bibfield  {author} {\bibinfo {author} {\bibfnamefont {I.}~\bibnamefont
  {Foulger}}, \bibinfo {author} {\bibfnamefont {S.}~\bibnamefont {Gnutzmann}},
  \ and\ \bibinfo {author} {\bibfnamefont {G.}~\bibnamefont {Tanner}},\
  }\bibfield  {title} {\enquote {\bibinfo {title} {Quantum search on graphene
  lattices},}\ }\href {\doibase 10.1103/PhysRevLett.112.070504} {\bibfield
  {journal} {\bibinfo  {journal} {Phys. Rev. Lett.}\ }\textbf {\bibinfo
  {volume} {112}},\ \bibinfo {pages} {070504} (\bibinfo {year}
  {2014})}\BibitemShut {NoStop}%
\bibitem [{\citenamefont {Tulsi}(2008)}]{tulsi08}%
  \BibitemOpen
  \bibfield  {author} {\bibinfo {author} {\bibfnamefont {A.}~\bibnamefont
  {Tulsi}},\ }\bibfield  {title} {\enquote {\bibinfo {title} {Faster
  quantum-walk algorithm for the two-dimensional spatial search},}\ }\href
  {\doibase 10.1103/PhysRevA.78.012310} {\bibfield  {journal} {\bibinfo
  {journal} {Phys. Rev. A}\ }\textbf {\bibinfo {volume} {78}},\ \bibinfo
  {pages} {012310} (\bibinfo {year} {2008})}\BibitemShut {NoStop}%
\bibitem [{\citenamefont {Abal}\ \emph {et~al.}(2012)\citenamefont {Abal},
  \citenamefont {Donangelo}, \citenamefont {Forets},\ and\ \citenamefont
  {Portugal}}]{abal12}%
  \BibitemOpen
  \bibfield  {author} {\bibinfo {author} {\bibfnamefont {G.}~\bibnamefont
  {Abal}}, \bibinfo {author} {\bibfnamefont {R.}~\bibnamefont {Donangelo}},
  \bibinfo {author} {\bibfnamefont {M.}~\bibnamefont {Forets}}, \ and\ \bibinfo
  {author} {\bibfnamefont {R.}~\bibnamefont {Portugal}},\ }\bibfield  {title}
  {\enquote {\bibinfo {title} {Spatial quantum search in a triangular
  network},}\ }\href {\doibase 10.1017/S0960129511000600} {\bibfield  {journal}
  {\bibinfo  {journal} {Math. Struct. Comput. Sci.%Mathematical Structures in Computer Science
  }\ }\textbf
  {\bibinfo {volume} {22}},\ \bibinfo {pages} {521--531} (\bibinfo {year}
  {2012})}\BibitemShut {NoStop}%
\bibitem [{\citenamefont {Manouchehri}\ and\ \citenamefont
  {Wang}(2006)}]{manouchehri2006}%
  \BibitemOpen
  \bibfield  {author} {\bibinfo {author} {\bibfnamefont {K.}~\bibnamefont
  {Manouchehri}}\ and\ \bibinfo {author} {\bibfnamefont {J.~B.}\ \bibnamefont
  {Wang}},\ }\href@noop {} {\emph {\bibinfo {title} {Physical implementation of
  quantum random walks}}},\ \bibinfo {type} {Tech. Rep.}\ (\bibinfo {year}
  {2006})\BibitemShut {NoStop}%
\bibitem [{\citenamefont {Gr{\"a}fe}\ \emph {et~al.}(2016)\citenamefont
  {Gr{\"a}fe}, \citenamefont {Heilmann}, \citenamefont {Lebugle}, \citenamefont
  {Guzman-Silva}, \citenamefont {Perez-Leija},\ and\ \citenamefont
  {Szameit}}]{Gr_fe_2016}%
  \BibitemOpen
  \bibfield  {author} {\bibinfo {author} {\bibfnamefont {M.}~\bibnamefont
  {Gr{\"a}fe}}, \bibinfo {author} {\bibfnamefont {R.}~\bibnamefont {Heilmann}},
  \bibinfo {author} {\bibfnamefont {M.}~\bibnamefont {Lebugle}}, \bibinfo
  {author} {\bibfnamefont {D.}~\bibnamefont {Guzman-Silva}}, \bibinfo {author}
  {\bibfnamefont {A.}~\bibnamefont {Perez-Leija}}, \ and\ \bibinfo {author}
  {\bibfnamefont {A.}~\bibnamefont {Szameit}},\ }\bibfield  {title} {\enquote
  {\bibinfo {title} {Integrated photonic quantum walks},}\ }\href {\doibase
  10.1088/2040-8978/18/10/103002} {\bibfield  {journal} {\bibinfo  {journal}
  {J. Opt.}\ }\textbf {\bibinfo {volume} {18}},\ \bibinfo {pages}
  {103002} (\bibinfo {year} {2016})}\BibitemShut {NoStop}%
\bibitem [{\citenamefont {B\"ohm}\ \emph {et~al.}(2015)\citenamefont {B\"ohm},
  \citenamefont {Bellec}, \citenamefont {Mortessagne}, \citenamefont {Kuhl},
  \citenamefont {Barkhofen}, \citenamefont {Gehler}, \citenamefont
  {St\"ockmann}, \citenamefont {Foulger}, \citenamefont {Gnutzmann},\ and\
  \citenamefont {Tanner}}]{bohm15}%
  \BibitemOpen
  \bibfield  {author} {\bibinfo {author} {\bibfnamefont {J.}~\bibnamefont
  {B\"ohm}}, \bibinfo {author} {\bibfnamefont {M.}~\bibnamefont {Bellec}},
  \bibinfo {author} {\bibfnamefont {F.}~\bibnamefont {Mortessagne}}, \bibinfo
  {author} {\bibfnamefont {U.}~\bibnamefont {Kuhl}}, \bibinfo {author}
  {\bibfnamefont {S.}~\bibnamefont {Barkhofen}}, \bibinfo {author}
  {\bibfnamefont {S.}~\bibnamefont {Gehler}}, \bibinfo {author} {\bibfnamefont
  {H.-J.}\ \bibnamefont {St\"ockmann}}, \bibinfo {author} {\bibfnamefont
  {I.}~\bibnamefont {Foulger}}, \bibinfo {author} {\bibfnamefont
  {S.}~\bibnamefont {Gnutzmann}}, \ and\ \bibinfo {author} {\bibfnamefont
  {G.}~\bibnamefont {Tanner}},\ }\bibfield  {title} {\enquote {\bibinfo {title}
  {Microwave experiments simulating quantum search and directed transport in
  artificial graphene},}\ }\href {\doibase 10.1103/PhysRevLett.114.110501}
  {\bibfield  {journal} {\bibinfo  {journal} {Phys. Rev. Lett.}\ }\textbf
  {\bibinfo {volume} {114}},\ \bibinfo {pages} {110501} (\bibinfo {year}
  {2015})}\BibitemShut {NoStop}%
\bibitem [{\citenamefont {Qiang~et al.}(2021)}]{Qiang21}%
  \BibitemOpen
  \bibfield  {author} {\bibinfo {author} {\bibfnamefont {X.}~\bibnamefont
  {Qiang}}, \bibinfo {author} {\bibfnamefont {Y.}~\bibnamefont
  {Wang}}, \bibinfo {author} {\bibfnamefont {S.}~\bibnamefont
  {Xue}}, \bibinfo {author} {\bibfnamefont {R.}~\bibnamefont
  {Ge}}, \bibinfo {author} {\bibfnamefont {L.}~\bibnamefont
  {Chen}}, \bibinfo {author} {\bibfnamefont {Y.}~\bibnamefont
  {Liu}}, \bibinfo {author} {\bibfnamefont {A.}~\bibnamefont
  {Huang}}, \bibinfo {author} {\bibfnamefont {X.}~\bibnamefont
  {Fu}}, \bibinfo {author} {\bibfnamefont {P.}~\bibnamefont
  {Xu}} \bibinfo {author} {\bibfnamefont {T.}~\bibnamefont
  {Yi}},  \emph {et~al.},\ }\bibfield  {title} {\enquote {\bibinfo {title}
  {Implementing graph-theoretic quantum algorithms on a silicon photonic
  quantum walk processor},}\ }\href {\doibase 10.1126/sciadv.abb8375}
  {\bibfield  {journal} {\bibinfo  {journal} {Sci. Adv.}\ }\textbf
  {\bibinfo {volume} {7}},\ \bibinfo {pages} {eabb8375} (\bibinfo {year} {2021})%,\ 10.1126/sciadv.abb8375
  }\BibitemShut {NoStop}%
\bibitem [{\citenamefont {Szameit}\ \emph {et~al.}(2006)\citenamefont
  {Szameit}, \citenamefont {Bl{\"o}mer}, \citenamefont {Burghoff},
  \citenamefont {Pertsch}, \citenamefont {Nolte},\ and\ \citenamefont
  {T{\"u}nnermann}}]{szameit2006}%
  \BibitemOpen
  \bibfield  {author} {\bibinfo {author} {\bibfnamefont {A.}~\bibnamefont
  {Szameit}}, \bibinfo {author} {\bibfnamefont {D.}~\bibnamefont {Bl{\"o}mer}},
  \bibinfo {author} {\bibfnamefont {J.}~\bibnamefont {Burghoff}}, \bibinfo
  {author} {\bibfnamefont {T.}~\bibnamefont {Pertsch}}, \bibinfo {author}
  {\bibfnamefont {S.}~\bibnamefont {Nolte}}, \ and\ \bibinfo {author}
  {\bibfnamefont {A.}~\bibnamefont {T{\"u}nnermann}},\ }\bibfield  {title}
  {\enquote {\bibinfo {title} {Hexagonal waveguide arrays written with fs-laser
  pulses},}\ }\href {\doibase https://doi.org/10.1007/s00340-005-2127-4}
  {\bibfield  {journal} {\bibinfo  {journal} {Appl. Phys. B}\ }\textbf
  {\bibinfo {volume} {82}},\ \bibinfo {pages} {507--512} (\bibinfo {year}
  {2006})}\BibitemShut {NoStop}%
\bibitem [{\citenamefont {Crespi}\ \emph {et~al.}(2012)\citenamefont {Crespi},
  \citenamefont {Longhi},\ and\ \citenamefont {Osellame}}]{crespi2012}%
  \BibitemOpen
  \bibfield  {author} {\bibinfo {author} {\bibfnamefont {A.}~\bibnamefont
  {Crespi}}, \bibinfo {author} {\bibfnamefont {S.}~\bibnamefont {Longhi}}, \
  and\ \bibinfo {author} {\bibfnamefont {R.}~\bibnamefont {Osellame}},\
  }\bibfield  {title} {\enquote {\bibinfo {title} {Photonic realization of the
  quantum rabi model},}\ }\href {\doibase 10.1103/PhysRevLett.108.163601}
  {\bibfield  {journal} {\bibinfo  {journal} {Phys. Rev. Lett.}\ }\textbf
  {\bibinfo {volume} {108}},\ \bibinfo {pages} {163601} (\bibinfo {year}
  {2012})}\BibitemShut {NoStop}%
\bibitem [{\citenamefont {Feng}\ \emph {et~al.}(2016)\citenamefont {Feng},
  \citenamefont {Wu}, \citenamefont {Zhao}, \citenamefont {Gao}, \citenamefont
  {Qiao}, \citenamefont {Yang}, \citenamefont {Lin},\ and\ \citenamefont
  {Jin}}]{feng2016}%
  \BibitemOpen
  \bibfield  {author} {\bibinfo {author} {\bibfnamefont {Z.}~\bibnamefont
  {Feng}}, \bibinfo {author} {\bibfnamefont {B.-H.}\ \bibnamefont {Wu}},
  \bibinfo {author} {\bibfnamefont {Y.-X.}\ \bibnamefont {Zhao}}, \bibinfo
  {author} {\bibfnamefont {J.}~\bibnamefont {Gao}}, \bibinfo {author}
  {\bibfnamefont {L.-F.}\ \bibnamefont {Qiao}}, \bibinfo {author}
  {\bibfnamefont {A.-L.}\ \bibnamefont {Yang}}, \bibinfo {author}
  {\bibfnamefont {X.-F.}\ \bibnamefont {Lin}}, \ and\ \bibinfo {author}
  {\bibfnamefont {X.-M.}\ \bibnamefont {Jin}},\ }\bibfield  {title} {\enquote
  {\bibinfo {title} {Invisibility cloak printed on a photonic chip},}\ }\href
  {\doibase https://doi.org/10.1038/srep28527} {\bibfield  {journal} {\bibinfo
  {journal} {Sci. Rep.}\ }\textbf {\bibinfo {volume} {6}},\ \bibinfo {pages}
  {1--8} (\bibinfo {year} {2016})}\BibitemShut {NoStop}%
\bibitem [{\citenamefont {Perets}\ \emph {et~al.}(2008)\citenamefont {Perets},
  \citenamefont {Lahini}, \citenamefont {Pozzi}, \citenamefont {Sorel},
  \citenamefont {Morandotti},\ and\ \citenamefont {Silberberg}}]{perets2008}%
  \BibitemOpen
  \bibfield  {author} {\bibinfo {author} {\bibfnamefont {H.~B.}\ \bibnamefont
  {Perets}}, \bibinfo {author} {\bibfnamefont {Y.}~\bibnamefont {Lahini}},
  \bibinfo {author} {\bibfnamefont {F.}~\bibnamefont {Pozzi}}, \bibinfo
  {author} {\bibfnamefont {M.}~\bibnamefont {Sorel}}, \bibinfo {author}
  {\bibfnamefont {R.}~\bibnamefont {Morandotti}}, \ and\ \bibinfo {author}
  {\bibfnamefont {Y.}~\bibnamefont {Silberberg}},\ }\bibfield  {title}
  {\enquote {\bibinfo {title} {Realization of quantum walks with negligible
  decoherence in waveguide lattices},}\ }\href {\doibase
  10.1103/PhysRevLett.100.170506} {\bibfield  {journal} {\bibinfo  {journal}
  {Phys. Rev. Lett.}\ }\textbf {\bibinfo {volume} {100}},\ \bibinfo {pages}
  {170506} (\bibinfo {year} {2008})}\BibitemShut {NoStop}%
\bibitem [{\citenamefont {Poulios}\ \emph {et~al.}(2014)\citenamefont
  {Poulios}, \citenamefont {Keil}, \citenamefont {Fry}, \citenamefont
  {Meinecke}, \citenamefont {Matthews}, \citenamefont {Politi}, \citenamefont
  {Lobino}, \citenamefont {Gr{\"a}fe}, \citenamefont {Heinrich}, \citenamefont
  {Nolte} \emph {et~al.}}]{poulios2014}%
  \BibitemOpen
  \bibfield  {author} {\bibinfo {author} {\bibfnamefont {K.}~\bibnamefont
  {Poulios}}, \bibinfo {author} {\bibfnamefont {R.}~\bibnamefont {Keil}},
  \bibinfo {author} {\bibfnamefont {D.}~\bibnamefont {Fry}}, \bibinfo {author}
  {\bibfnamefont {J.~D.~A.}\ \bibnamefont {Meinecke}}, \bibinfo {author}
  {\bibfnamefont {J.~C.~F.}\ \bibnamefont {Matthews}}, \bibinfo {author}
  {\bibfnamefont {A.}~\bibnamefont {Politi}}, \bibinfo {author} {\bibfnamefont
  {M.}~\bibnamefont {Lobino}}, \bibinfo {author} {\bibfnamefont
  {M.}~\bibnamefont {Gr{\"a}fe}}, \bibinfo {author} {\bibfnamefont
  {M.}~\bibnamefont {Heinrich}}, \bibinfo {author} {\bibfnamefont
  {S.}~\bibnamefont {Nolte}},  \emph {et~al.},\ }\bibfield  {title} {\enquote
  {\bibinfo {title} {Quantum walks of correlated photon pairs in
  two-dimensional waveguide arrays},}\ }\href {\doibase
  10.1103/PhysRevLett.112.143604} {\bibfield  {journal} {\bibinfo  {journal}
  {Phys. Rev. Lett.}\ }\textbf {\bibinfo {volume} {112}},\ \bibinfo {pages}
  {143604} (\bibinfo {year} {2014})}\BibitemShut {NoStop}%
\bibitem [{\citenamefont {Caruso}\ \emph {et~al.}(2016)\citenamefont {Caruso},
  \citenamefont {Crespi}, \citenamefont {Ciriolo}, \citenamefont {Sciarrino},\
  and\ \citenamefont {Osellame}}]{caruso2016}%
  \BibitemOpen
  \bibfield  {author} {\bibinfo {author} {\bibfnamefont {F.}~\bibnamefont
  {Caruso}}, \bibinfo {author} {\bibfnamefont {A.}~\bibnamefont {Crespi}},
  \bibinfo {author} {\bibfnamefont {A.~G.}\ \bibnamefont {Ciriolo}}, \bibinfo
  {author} {\bibfnamefont {F.}~\bibnamefont {Sciarrino}}, \ and\ \bibinfo
  {author} {\bibfnamefont {R.}~\bibnamefont {Osellame}},\ }\bibfield  {title}
  {\enquote {\bibinfo {title} {Fast escape of a quantum walker from an
  integrated photonic maze},}\ }\href {\doibase
  https://doi.org/10.1038/ncomms11682} {\bibfield  {journal} {\bibinfo
  {journal} {Nat. Comm.}\ }\textbf {\bibinfo {volume} {7}},\ \bibinfo {pages}
  {1--7} (\bibinfo {year} {2016})}\BibitemShut {NoStop}%
\bibitem [{\citenamefont {Tang}\ \emph {et~al.}(2018)\citenamefont {Tang},
  \citenamefont {Di~Franco}, \citenamefont {Shi}, \citenamefont {He},
  \citenamefont {Feng}, \citenamefont {Gao}, \citenamefont {Sun}, \citenamefont
  {Li}, \citenamefont {Jiao}, \citenamefont {Wang} \emph {et~al.}}]{tang2018}%
  \BibitemOpen
  \bibfield  {author} {\bibinfo {author} {\bibfnamefont {H.}~\bibnamefont
  {Tang}}, \bibinfo {author} {\bibfnamefont {C.}~\bibnamefont {Di~Franco}},
  \bibinfo {author} {\bibfnamefont {Z.-Y.}\ \bibnamefont {Shi}}, \bibinfo
  {author} {\bibfnamefont {T.-S.}\ \bibnamefont {He}}, \bibinfo {author}
  {\bibfnamefont {Z.}~\bibnamefont {Feng}}, \bibinfo {author} {\bibfnamefont
  {J.}~\bibnamefont {Gao}}, \bibinfo {author} {\bibfnamefont {K.}~\bibnamefont
  {Sun}}, \bibinfo {author} {\bibfnamefont {Z.-M.}\ \bibnamefont {Li}},
  \bibinfo {author} {\bibfnamefont {Z.-Q.}\ \bibnamefont {Jiao}}, \bibinfo
  {author} {\bibfnamefont {T.-Y.}\ \bibnamefont {Wang}},  \emph {et~al.},\
  }\bibfield  {title} {\enquote {\bibinfo {title} {Experimental quantum fast
  hitting on hexagonal graphs},}\ }\href {\doibase 10.1038/s41566-018-0282-5}
  {\bibfield  {journal} {\bibinfo  {journal} {Nat. Photon.}\ }\textbf
  {\bibinfo {volume} {12}},\ \bibinfo {pages} {754--758} (\bibinfo {year}
  {2018})}\BibitemShut {NoStop}%
\end{thebibliography}
\end{document}